# Metasurface embodied intelligence through electromagnetic world model

Che Liu[1,†,*], Zhenhao Fu[1,†], Qian Ma[1], Jiajing Wu[1], Wen Ming Yu[1], Lianlin Li[2] and Tie Jun Cui[1,3,4,*]

[1]State Key Laboratory of Millimeter Waves, Southeast University, Nanjing 211189, China

[2]State Key Laboratory of Photonics and Communications, Peking University, Beijing 100871, China

[3]Institute of Electromagnetic Space, Southeast University, Nanjing 211189, China

[4]Center of Metamaterials, Suzhou National Laboratory, Suzhou 215164, China

*E-Mail: cheliu@seu.edu.cn, tjcui@seu.edu.cn

[†]Equally contributed to this work.

## ABSTRACT

Mastering invisible electromagnetic (EM) environment and sculpting radio waves with the dexterity of manipulating light or matter have long been aspirations in physics and information science. While information metasurfaces (IMSs) provide the physical interface to program EM wavefields, their real-world autonomy is fundamentally limited by environmental 'blindness' and the prohibitive overhead of site-specific and trial-and-error retraining. Here we propose metasurface embodied intelligence through world model (metaEI-WM), a universal and out-of-the-box paradigm that achieves expert-level performance without on-site fine-tuning. In contrast to purely data-driven agents, metaEI-WM establishes a fundamental understanding of the EM dynamics by integrating fully automated semantic environment modelling with embedded electrodynamic priors. By anticipating future scenarios *in silico*, it optimizes the IMS coding configurations to dynamically shape EM environments on demand. We show that metaEI-WM successfully enables zero-latency non-line-of-sight signal enhancements, symbiotic communications, and contactless physiological sensing across highly complex and unseen indoor scenarios. To the best of our knowledge, metaEI-WM is the first paradigm to achieve end-to-end automation of complex spatial channel manipulation tasks *ab initio*, requiring neither human-annotated data nor online training. This framework bridges the gap between digital intelligence and physical-layer wave dynamics, offering a scalable solution for robust and self-managing wireless ecosystems.

## Introduction

Since the inception of wireless technology, the electromagnetic (EM) space has predominantly been treated as a passive and uncontrollable propagation medium. The ultimate scientific pursuit has been to transform this chaotic and invisible realm into a fully programmable entity, sculpting the EM wave trajectories to bypass obstacles, focus energy, or synthesize new communication channels *ab initio*. Information metasurfaces (IMSs)[1-5] have recently emerged as physical hardware to realize this vision, standing at the forefront of sixth-generation (6G) wireless revolution[6-8] by translating digital command into physical-layer wave manipulation and sensing[9-16]. However, bridging the gap between hardware capabilities and true environmental autonomy remains a formidable challenge. The practical utility of most existing IMS systems is fundamentally limited by the need for either resource-intensive in situ channel estimation prior to deployment[17], or a reliance on human experts for online training to achieve optimal configurations[18,19]. Although recent advancements in IMS with integrated detectors have enabled the sensing of base station[9,20] and user positions[21], these reactive systems exhibit severe vulnerability in complex multi-path and multi-user environments or when user blockages cause signals to plummet below detection thresholds. While the previous metaAgent[22] successfully demonstrated autonomous operations, its real-world scalability is severely restricted by the necessity for weeks of site-specific training and an extreme sensitivity to environmental dynamics that triggers costly and prohibitive retraining.

Here, we propose metasurface embodied intelligence based on world model (metaEI-WM), an autonomous framework that pioneers the integration of world models into the invisible EM domain. Moving beyond the empirical trial-and-error, metaEI-WM approaches EM manipulations through the lens of a cyber-physical agent. By automatically transforming the unseen indoor environments into geometry- and material-aware digital twins, the system integrates spatial semantics with predictive electrodynamics to construct its internal world model. This allows metaEI-WM to anticipate the spatial evolution of radio waves and optimize the metasurface configurations *in silico* before any physical action is executed, enabling universal and 'out-of-the-box' deployment. Ultimately, it represents the first end-to-end realization of a self-managing intelligent radio ecosystem that completely eliminates the necessity for human-curated datasets and continuous online training protocols.

The operational architecture of metaEI-WM relies on the seamless convergence of three principal modules (Fig. 1) driven by a vision-language model (Extended Data Fig. 1 and Supplementary Note 1), which functions as the cognitive center processing natural language directives.

We empirically validate the robust capabilities of metaEI-WM across highly complex and distinct physical environments, including the typical workplaces, intricate multistory corridors, and heavily compartmentalized residential apartments. Across these diverse settings, the framework consistently delivers signal enhancements in non-line-of-sight dead zones that rivaled the human-expert on-site optimizations, achieving average power gains between 5.10 dB and 7.91 dB while reducing bit error rates by up to 1.11 orders of magnitude. Beyond these enhancements, we further demonstrate the system's capacity for active physical-layer signal synthesis (e.g. symbiotic communications) and physiological monitoring. Collectively, these empirical achievements underscore the system's exceptional proficiency in transforming the passive architectural confines into actively programmable and intelligent wireless ecosystems.

## Functional architecture

The metaEI-WM achieves its functionality through the synergy of three critical components (Fig. 1): a heterogeneous perception and representation module leverages the robotic assistance for semantic-level environmental modeling, an EM dynamics prediction module forecasts the spatial EM outcomes of prospective actions, and a decision-and-execution module selects and implements the optimal IMS coding configurations. This section introduces these components.

### Perception and representation module

The perception and representation module autonomously converts unseen indoor environments into spatial semantics of the actionable EM physical world model (Fig. 2), eliminating the need for manual surveying or site-specific retraining. Deployed on a smart mobile 'wingman' robot, metaEI-WM acquires four critical environmental priors: the poses of deployed IMSs, the spatial coordinates of radio sources, a geometry- and material-aware digital twin, and a semantic-topological understanding of the scene.

The metaEI-WM localizes IMSs and radio sources to establish the programmable boundaries and radiating entities of wireless environment. Exploiting the periodic visual textures of metasurfaces, the system recovers the exact identities and six-degree-of-freedom poses of IMSs via a vision-based homography decomposition[23] (see Methods section 'Automated localization of IMS'). Concurrently, it localizes the signal sources by applying the SpotFi principle[24] to channel state information (CSI) gathered during the autonomous traversal. By resolving multipath directions using 2D-MUSIC and intersecting line-of-sight bearing rays from multiple robot viewpoints, the source position is obtained as the point that best intersects the observed bearing rays (the least-squares ray-intersection estimator):

$$\mathbf{p}_s^* = \arg\min_{\mathbf{p}} \sum \left\| \left(\mathbf{I} - \mathbf{u}_k \mathbf{u}_k^T\right)(\mathbf{p} - \mathbf{r}_k) \right\|^2$$

where $\mathbf{r}_k$ is the robot position at the k-th observation and $\mathbf{u}_k$ is the estimated arrival direction (see Methods section 'Radio-source localization').

Meanwhile, metaEI-WM leverages modern foundation models[25-28] to reconstruct the surrounding indoor scene into spatial-semantic world model representation with a geometry- and material-aware digital twin. During the autonomous exploration, the TARE hierarchical framework[25] ensures efficient and blind-spot-free spatial coverage, while R3LIVE[26] concurrently fuses tightly coupled LiDAR-camera-IMU odometry to generate dense, globally registered and RGB-coloured point clouds. To extract the structural semantics, SpatialLM[27] directly parses these point clouds into a 'scene-as-code' format:

$$\mathcal{O} = \{(\mathbf{b}_i, s_i, \mathbf{a}_i)\}_{i=1}^{N}$$

composing of the oriented 3D bounding box $\mathbf{b}_i$, semantic label $s_i$, and spatial attributes $\mathbf{a}_i$ of each object (such as walls, doors, windows, furniture and appliances). This scene-as-code format allows seamless parsing by both electromagnetic simulator and decision agent, enabling new object categories or reflector types to be introduced through natural-language prompts. Complementing this object-level extraction, RoomFormer[28] projects the dense reconstruction into a bird's-eye view to predict the macroscopic room topology, defining the functional boundaries and inter-room spatial relationships. This hierarchical parsing elevates the representation from an unstructured point set to a semantically and topologically structured scene (Extended Data Fig. 2).

The metaEI-WM bridges the semantic-to-physical gap by mapping the extracted object semantic labels to standard EM material categories (e.g., concrete, glass, wood, metal, plastic, ceramic or fabric) based on the ITU-R P.2040 library. The frequency-dependent permittivity and conductivity are assigned as:

$$m_i = \xi(s_i)$$
$$\theta_i(f) = \{\varepsilon_{r,i}(f), \sigma_i(f)\}$$

where $s_i$ is the semantic label, $m_i$ is the matched material class, and $\theta_i(f)$ contains the relative permittivity and conductivity at frequency $f$.

The ultimate spatial semantics for world model representation is encapsulated as:

$$\mathcal{W}_{EM} = \{\mathcal{G}, \mathcal{O}, \mathcal{T}, \mathcal{M}_{EM}, \mathcal{Q}_{IMS}, \mathcal{P}_s\}$$

where $\mathcal{G}$ is the reconstructed geometry, $\mathcal{O}$ the semantic object set, $\mathcal{T}$ the room topology, $\mathcal{M}_{EM}$ the material assignment, $\mathcal{Q}_{IMS}$ the IMS poses, and $\mathcal{P}_s$ the signal-source positions. Consequently, metaEI-WM constructs not merely a visual map, but a comprehensive semantic-EM digital twin, in which metaEI-WM knows what each object is, what material is made of, which room belongs to and how different rooms are connected. This structured prior enables out-of-the-box EM prediction and IMS configuration optimization across structurally diverse environments without requiring prior training. Further implementation details of this section are provided in Supplementary Note 2, Supplementary Note 3 and Supplementary Video 1.

**EM dynamics prediction module**

Serving as the system's electrodynamic *imagination*, the EM dynamics prediction module leverages the spatial semantics to forecast the wave propagation prior to any physical metasurface actuation. Given the structural geometry, material properties, transceiver coordinates, and IMS poses, the module anticipates the spatial channel responses under arbitrary programmable configurations. For a user-specified target region $\Omega$, the prediction module[29] resolves the dominant line-of-sight (LoS), environmental multipath, and IMS-cascaded trajectories, translating high-level semantic instructions into a physically grounded ray map:

$$\mathcal{R}_\Omega = \chi(\mathcal{W}_{\mathrm{EM}}, f, \Omega)$$

where $f$ is the carrier frequency and $\chi(\cdot)$ represents the ray-tracing operator. Unlike the purely statistical channel models, this deterministic approach preserves rigorous causal links between the architectural semantics (e.g., walls, furniture, diffracting edges), metasurface states, and the resulting EM field distribution.

The synthesized channel at any spatial coordinate $\mathbf{r}$ is formulated as a coherent superposition of three physically interpretable mechanisms:

$$h(\mathbf{r},\mathbf{\Phi}) = h_{dir}(\mathbf{r}) + h_{env}(\mathbf{r}) + \mathbf{g}_{\mathrm{IMS}\rightarrow\mathbf{r}}^{T}\mathbf{\Phi}\mathbf{f}_{\mathrm{Tx}\rightarrow\mathrm{IMS}}$$

where $h_{dir}$ and $h_{env}$ indicate the direct and environment-induced multipath components, respectively. The final term captures the programmable cascaded channel, where $\mathrm{f}_{\mathrm{Tx}\rightarrow\mathrm{IMS}}$ and $g_{\mathrm{IMS}\rightarrow\mathbf{r}}^{T}$ denote the incident and re-radiated channel vectors on the metasurface aperture, and $\mathbf{\Phi}$ represents the IMS coding matrix. Consequently, the predicted received power is derived as:

$$P(\mathbf{r},\mathbf{\Phi}) = P_{Tx}|h(\mathbf{r},\mathbf{\Phi})|^{2}$$

This compact analytical formulation empowers metaEI-WM to evaluate the consequences of candidate IMS configurations *in silico*, circumventing the prohibitive overhead of exhaustive online probing.

Operationally (Fig. 3b), this predictive capability is engaged once a user's semantic command (e.g., 'enhance the second bedroom') is mapped to a specific region in the world model. By simulating the multi-bounce interactions among the transmitter, environment, and IMS, the module yields a region-aware estimation of the radio coverage. By supplying the subsequent decision-and-execution module with a searchable physical landscape, this prediction framework fundamentally transforms the metasurface control from empirical trial-and-error into deterministic and model-based EM planning (see Supplementary Note 4 for detailed deterministic channel construction).

**Decision-and-execution module**

The decision-and-execution module translates the predicted EM dynamics into hardware-executable IMS coding configurations. Having grounded the semantic target in the world model, metaEI-WM bypasses the empirical trial-and-error in the physical domain. Instead, it formulates the user-specified target region $\Omega$ as a deterministic optimization objective. The module firstly extracts a compact subset of effective dominant rays from the prediction engine:

$$\mathcal{L}_{\Omega} = \{\mathbf{r}_{m}\}_{m=1}^{M}$$

where each ray $\mathbf{r}_{m}$ encapsulates the geometric trajectory, attenuation, and phase accumulation linking the transmitter, IMS aperture, and target area. This filtered ray map constitutes a physically interpretable search space that deliberately discards the negligible paths. The control objective is thus abstracted as finding an optimal IMS coding matrix $\mathbf{\Phi}^*$ that maximizes a task-specific utility $\mathcal{U}$ (e.g., regional received power or signal quality) in the hardware-constrained discrete coding space $\mathbb{Q}$:

$$\mathbf{\Phi}^* = \arg\max_{\mathbf{\Phi}\in\mathbb{Q}} \mathcal{U}(\Omega, \mathbf{\Phi}|\mathcal{L}_\Omega)$$

To solve the non-convex problem efficiently without exhaustive enumeration, metaEI-WM employs a physics-informed prediction-inversion strategy. Exploiting EM time-reversal symmetry, the target blind region is conceptually modeled as a virtual source. The desired field is back-projected along the dominant ray paths onto the IMS aperture, establishing an idealized radiation pattern for effective target illumination. A modified Gerchberg-Saxton procedure[30] then translates this idealized aperture field into a realizable reflection phase distribution. Analytically, the required phase shift at the aperture coordinate **x** is:

$$\Delta\varphi(\mathbf{x}) = \varphi_{tar}(\mathbf{x}) - \varphi_{inc}(\mathbf{x})$$

where $\varphi_{tar}$ is the phase of the back-projected target field, and $\varphi_{inc}$ is the incident phase derived from the ray-tracing engine. The continuous phase profile is subsequently discretized into a permissible hardware alphabet:

$$\mathcal{C}(\mathbf{x}) = Q[\Delta\varphi(\mathbf{x})]$$

where $Q[\cdot]$ denotes the quantization operator (e.g., binary quantization for the 1-bit metasurfaces used in our experiments).

Figure 3c illustrates this staged optimization process of IMS configuration in decision pipeline. Following the spatial ray filtering and time-reversal back-projection, the Gerchberg-Saxton inversion decouples the reflection phase from the incident wave, yielding the discrete coding matrix. Crucially, metaEI-WM performs the forward field synthesis to verify this action *in silico* before physical execution. Furthermore, the framework remains fundamentally agnostic to the underlying metasurface hardware. Provided that the physical device parameters (operating frequency, aperture geometry, unit-cell lattice, and phase states) are registered in the world model, the system will output universally compatible control signals, as demonstrated by executing the same pipeline across three heterogeneous IMS platforms (Fig. 3a).

The physical execution step also closes the loop between prediction and reality. Should real-time measurements indicate that the achieved signal quality falls significantly below the expected threshold (often due to unmodelled human blockages, furniture displacements, or residual modelling errors), metaEI-WM triggers a lightweight local refinement around the predicted optimum. The converged optimal coding pattern, paired with its spatial coordinates and semantic context, is permanently embedded into a spatiotemporal knowledge graph. Consequently, when encountering analogous topologies or semantic instructions, the system retrieves previously validated configurations rather than computing them *ab initio*. This memory mechanism elevates metaEI-WM from a one-time optimizer to a self-improving and embodied cyber-physical agent capable of accumulating reusable EM experience (see Supplementary Note 5 and Supplementary Video 2).

## Evaluation

To validate the practical efficacy of the metaEI-WM architecture and its capacity to translate in *silico* EM reasoning into physical reality, we systematically deployed the framework across a spectrum of highly unstructured indoor environments. Moving beyond the theoretical formulation of the digital twin, the following empirical evaluations demonstrate the system's out-of-the-box adaptability in overcoming severe real-world propagation complexities. Here, we quantify its performance across three distinct operational frontiers: autonomous dead-zone signal enhancement, symbiotic ambient communications, and contactless physiological sensing, collectively confirming its ability to transform the passive spatial landscapes into intelligently programmable wireless ecosystems.

### Benchmark scenarios

We deployed the system across three unstructured scenarios to confront the profound EM complexities, ranging from severe multipath scattering to deep non-line-of-sight fading (see Methods for hardware details).

- **Workplace.** Validate the predictive channel reconstructions under the typical indoor multipath scattering, driven by the diverse absorption and refraction signatures of standard structural materials (e.g., wood, glass, and reinforced concrete).
- **Multistory corridor.** Extend the validation from single-level environments to multi-story 3D structures. The presence of stepped staircases dictates that wave propagation is heavily governed by edge diffraction and non-coplanar Fresnel reflections, demanding superior fidelity from the 3D environmental modelling and prediction modules.
- **Residential apartment.** Test the ultimate translational potentials in a highly compartmentalized

dwelling. Rich semantic landscape and extreme material heterogeneity (spanning from absorptive sofas to highly reflective appliances) induce profound random scattering, rigorously challenging the architecture's analytical resilience.

**Dead zone signal enhancement**

We assessed the system's core capability to recover deep non-line-of-sight (NLOS) dead zones. Across all three scenarios (Fig. 4, Extended Data Figs. 3 and 4), the framework autonomously improved the received localized power by 5.1-7.91 dB and reduced the bit error rate (BER) by 0.8-1.11 orders of magnitude, entirely without the on-site retraining.

Specifically, in the workplace (Fig. 4a and Extended Data Fig. 5a), empirical measurements across 60 spatial points validated the world model's high predictive fidelity, with Willmott Index of Agreement (IOA)[31] values of 0.9316 and 0.9405 before and after the modulation, respectively (see Extended Data Table 1). The physical intervention yielded an average measured power gain of 6.47 dB and a 0.84 order-of-magnitude BER reduction. In the corridor (Fig. 4b and Extended Data Fig. 5b), where the propagation is heavily governed by complex stair-edge diffraction, the system maintained the strong predictive correlation (IOA: 0.9205 before, 0.8366 after), achieving a 7.91 dB gain and a 0.80 order-of-magnitude BER reduction.

In the compartmentalized residential apartment, a deep NLOS dead zone (the second bedroom) obstructed by walls and a staggered corridor was selected via natural-language commands. Guided by the semantic-topological map, metaEI-WM autonomously distributed 20 optimized spatial nodes in the target region to ensure comprehensive coverage of the entire room (Fig. 4c). The world model maintained high predictive fidelity (Fig. 4f). Physically, the optimized IMS yielded an *in situ* average power gain of 5.1 dB (closely tracking the 5.99 dB prediction) (Fig. 4i), driving the most substantial link-level recovery across all tested scenarios: a 1.11 order-of-magnitude reduction in bit error rate (BER) (Fig. 4l). This was further corroborated by the highly compact I/Q constellations, indicating superior modulation-domain separability (Extended Data Fig. 5c). Beyond raw physical metrics, we validated application-level robustness via a 30-s video transmission. The optimized channel practically eliminated the mosaic artifacts and freezing frames during video transmission, confirming that metaEI-WM effectively translates the spatial power focusing into stable and high-quality video delivery (see Extended Data Fig. 5d, Supplementary Video 3).

### Symbiotic communications

Beyond recovering weak links, metaEI-WM can reshape ambient waves into a multiplexed wireless medium (the mathematical modeling descriptions are available in the Methods section 'Symbiotic communication'). We demonstrated a symbiotic communication paradigm in the corridor setup (Fig. 5a), where the metasurface simultaneously served as a primary Wi-Fi user and a secondary IoT receiver. Guided by the world model, the system autonomously generated a dual-state coding strategy (Fig. 5b). By satisfying the phase constraint $\angle H_1(\mathbf{\Phi}_1) \approx \angle H_1(\mathbf{\Phi}_0)$, the metasurface preserved the phase stability for the primary phase-shift-keying link. Concurrently, it produced distinguishable field amplitudes toward the secondary receiver, successfully superimposing an amplitude-shift-keying modulated image stream at approximately 1 Kbps without requiring a dedicated radio-frequency source (Fig. 5c).

### Contactless physiological sensing

The dynamic spatial focusing capability of metaEI-WM further enables the system to function as a contactless sensor for mobile users (see Methods section and Supplementary Note 6 for details). During mobility tests, the system accurately tracked the user trajectories to continuously optimize the IMS configurations, thereby directly enhancing the empirical channel capacity of handheld devices (Extended Data Fig. 6). For respiratory monitoring, the system transitioned to human skeletal tracking, identifying the chest-spine keypoint as the EM focusing ancho and encoding the subtle respiratory displacements as the phase modulations within the backscattered echo (Fig. 5d). Utilizing a variational mode decomposition-based pipeline to filter the environmental clutter and body sway[32], metaEI-WM successfully extracted a dynamic respiration-rate curve that closely matched to a wearable reference device (Fig .5e-f and Supplementary Video 4).

## Conclusion

We presented metaEI-WM, a universal cyber-physical paradigm that masters the complex spatial EM wave manipulation without requiring site-specific fine-tuning. By illustrating exceptional resilience against extreme multipath fading and severe non-line-of-sight blockages across highly unstructured indoor environments, this framework fundamentally accelerates the scalable real-world deployment of programmable wireless ecosystems. To the best of our knowledge, it represents the first out-of-the-box architecture to autonomously execute zero-latency signal recovery alongside wireless symbiotic communications and contactless physiological sensing *ab initio*. We hope that the approach contributes a meaningful building block to the ongoing evolution of intelligent radio environments.

Importantly, the mobile perception platform (wingman robot) completely bypasses the need for bespoke hardware. Its unified algorithm stack is designed for seamless integration into conventional off-the-shelf service robots or commercial robot vacuums provided that they are equipped with the standard depth, inertial, and visual sensors. Grounded in an electrodynamic world model, metaEI-WM unlocks new trajectories for future research. This foundational architecture enables artificial agents to extract the generalized EM knowledge from diverse architectural layouts and learn the universal foundation models across varying radio domains. Consequently, it provides a robust pathway for intelligent EM agents to accumulate the generalized competency required to sustain next-generation wireless networks.

## METHODS

### Experimental scenarios and hardware configurations

**Workplace.** The environment comprises of office and warehouse regions featuring heterogeneous materials (e.g., wood, glass, reinforced concrete). Following the autonomous robotic reconstruction of scene's semantic topology, the office was divided into five sub-areas. A 10-dBm omnidirectional transmitter and a 3.5-GHz 1-bit programmable reflective metasurface were deployed in the warehouse, with all phase and geometric centers aligned horizontally at a height of 1.5 m. To evaluate the regional enhancement, 60 spatial sampling points (12 per sub-area) were probed using a high-gain directional Vivaldi receiving antenna. Additionally, dynamic user tracking and contactless respiration monitoring were conducted in Area 1 and the warehouse, respectively.

**Multistory corridor.** Featuring concrete walls, marble floors, and structural staircases, this setup co-located a 10-dBm omnidirectional transmitter and a 5-GHz programmable reflective metasurface in the main corridor, both centred at 1.7 m above the floor. Targeting the lower staircase, 20 sampling points were uniformly distributed along ascending steps. An omnidirectional receiver was positioned at each point, maintaining a relative phase-centre height of 1.25 m above the local supporting plane (floor or stair tread). This specific geometry was also utilized for the symbiotic communication test to simultaneously serve a primary PSK link and a secondary ASK-modulated backscatter stream.

**Residential apartment.** Conducted in a compartmentalized dwelling, this setup placed a 10-dBm omnidirectional transmitter and a 5-GHz magnetically actuated flipping metasurface[33] (44 cm × 48 cm aperture) at the entryway. To mimic typical household-router placement, both devices were fixed at 0.75 m height. The target was a severe non-line-of-sight second bedroom obstructed by walls and a staggered corridor. Guided by the semantic map, metaEI-WM automatically generates 20 validation points within this target room. Measurements (received power, BER, and video transmission) were conducted at a 1.5-m observation plane to approximate the typical holding height of mobile devices.

### Communication and video-quality measurements

Received power is reported in dBm. For each scenario, the localized power gain was computed as the difference between the optimized and baseline states at identical spatial coordinates:

$$G^i = P^i_{with} - P^i_{base}$$

Scenario-level values are reported as $mean \pm standard\ deviation\ (s.d.)$ across the respective spatial sampling points. The predictive fidelity of the digital twin was quantitatively evaluated using the Willmott index of agreement:

$$d = 1 - \frac{\sum_{i=1}^{n} \left(P^i_{sim} - P^i_{meas}\right)^2}{\sum_{i=1}^{n} \left(\left|P^i_{sim} - \overline{P}_{meas}\right| + \left|P^i_{meas} - \overline{P}_{meas}\right|\right)^2}$$

where $P^i_{sim}$ and $P^i_{meas}$ denote the simulated and measured received power, respectively, and $\overline{P}_{meas}$ is the mean measured power.

Bit error rate (BER) and I/Q constellations were derived from synchronized baseband samples captured before and after metasurface optimization, following standard carrier and timing recovery procedures. To evaluate application-level robustness, a 30-s transport-stream video was transmitted over the residential wireless link. The received video quality was objectively quantified using the structural similarity index measure (SSIM)[34] and a custom smoothness score:

$$Q = 100(1 - \alpha F - \beta B)$$

where $F$ represents the temporal-freezing penalty (derived from inter-frame variation mismatches), $B$ denotes the normalized block-artifact penalty (derived from abnormal intensity discontinuities along block boundaries), and $\alpha$ and $\beta$ are empirical weighting coefficients.

**Automated localization of IMS**

To register each IMS as a programmable boundary within the digital twin, its 6-degree-of-freedom (6-DoF) pose is recovered via vision-based template matching. Using observations from the wingman robot or a fixed ZED 2i camera, ORB features[35] are extracted from grayscale frames and matched against pre-stored orthogonal-view templates. To mitigate false correspondences induced by periodic metasurface unit cells, a nearest-neighbour ratio test and RANSAC-based geometric verification[36] are sequentially applied to estimate a robust homography matrix $\mathbf{H}$, with the highest-inlier template determining the IMS identity.

Assuming a calibrated camera intrinsic matrix $\mathbf{K}$, the planar projection of a metasurface point $\mathbf{q} = [X, Y, 1]^T$ to its homogeneous image coordinate $\mathbf{x} = [u, v, 1]^T$ is formulated as:

$$s\mathbf{x} = \mathbf{H}\mathbf{q} = \mathbf{K}[\mathbf{r}_1, \mathbf{r}_2, \mathbf{t}]\mathbf{q}$$

where $s$ is a scale factor. Letting $\mathbf{H} = [\mathbf{h}_1, \mathbf{h}_2, \mathbf{h}_3]$, translation vector $\mathbf{t}$ and initial rotation columns $\mathbf{r}_1, \mathbf{r}_2$ are analytically decomposed as:

$$\lambda_h = \frac{2}{\|\mathbf{K}^{-1}\mathbf{h}_1\| + \|\mathbf{K}^{-1}\mathbf{h}_2\|}$$

$$\mathbf{r}_1 = \lambda_h\, \mathbf{K}^{-1}\mathbf{h}_1,\ \mathbf{r}_2 = \lambda_h\, \mathbf{K}^{-1}\mathbf{h}_2,\ \mathbf{t} = \lambda_h\, \mathbf{K}^{-1}\mathbf{h}_3,\ \mathbf{r}_3 = \mathbf{r}_1 \times \mathbf{r}_2$$

To strictly enforce orthogonality against image noise and imperfect feature localization, the initial rotation matrix $\widehat{\mathbf{R}} = [\mathbf{r}_1, \boldsymbol{r}_2, \mathbf{r}_3]$ is projected onto the $SO(3)$ manifold via singular value ($\widehat{\mathbf{R}} = \mathbf{U\Sigma V}^{\mathrm{T}}$). The corrected rotation matrix is computed as:

$$\mathbf{R} = \mathbf{U}\mathrm{diag}\left(1{,}1, \det\left(\mathbf{U}\mathbf{V}^T\right)\right)\mathbf{V}^T$$

This procedure yields a rigid and geometrically verified 6-DoF pose (centre coordinates and yaw-pitch-roll orientation), which is subsequently passed to the digital twin to spatially anchor each programmable boundary condition.

**Radio-source localization**

Radio transmitters are localized using channel state information (CSI) acquired by the wingman robot during autonomous traversal (Intel Wi-Fi Link 5300, Ubuntu 14.04, Linux 802.11n CSI Tool). Operating at a 20-MHz bandwidth, the system records the complex frequency response across 3 antennas and 30 subcarriers. The measured CSI for the $m$-th antenna and $k$-th subcarrier is modelled as a superposition of $L$ multipath components (as the SpotFi implementation described in the Supplementary Information):

$$H_{m,k} = \sum_{p=1}^{L} \gamma_p e^{-j2\pi f_k \tau_p} e^{-j2\pi d(m-1)\sin(\theta_p)/\lambda} + \eta_{m,k}$$

where $\gamma_p$, $\tau_p$ and, $\theta_p$ denote the complex path gain, time of flight (ToF), and angle of arrival (AoA) of the $p$-th component, respectively. Here, $f_k$ is the subcarrier frequency, $d$ the antenna spacing, $\lambda$ the carrier wavelength, and $\eta_{m,k}$ the measurement noise.

Following packet-level phase sanitization to mitigate sampling-frequency and timing offsets, the 3 × 30 CSI matrix undergoes spatial-frequency smoothing to construct an augmented measurement matrix $\mathbf{Y}$ containing $Q$ smoothed snapshots. The noise subspace $\mathbf{E}_N$ is isolated via eigenvalue decomposition of the sample covariance matrix $\mathbf{R}_y = \frac{1}{Q}\mathbf{Y}\mathbf{Y}^H$. A joint 2D-MUSIC pseudo-spectrum is then evaluated:

$$P(\theta, \tau) = \frac{1}{\mathbf{a}(\theta, \tau)^H \mathbf{E}_N \mathbf{E}_N^H \mathbf{a}(\theta, \tau)}$$

where $\mathbf{a}(\theta, \tau)$ is the joint steering vector, and $(\cdot)^H$ denotes the Hermitian transpose.

To circumvent absolute ToF ambiguities induced by hardware desynchronization, candidate AoA–ToF peaks are clustered across consecutive packets. Adopting the SpotFi principle, the line-of-sight (direct-path) AoA is robustly identified based on the temporal support and angular-delay consistency of the clusters. Finally, the selected direct-path bearing rays from multiple known robot poses are fused via the least-squares intersection estimator to determine the 3D spatial coordinates of the transmitter. These coordinates are subsequently registered as fixed radiating sources within the physical-electromagnetic world model.

**Symbiotic communication**

Symbiotic communication[37] repurposes the ambient Wi-Fi carrier $x_p(t)$ as both an energy source and a substrate for secondary information, eliminating the need for an independent radio-frequency transmitter. In this shared propagation system, the baseband signal at receiver $k \in \{1,2\}$ is modeled as:

$$y_k(t,b) = \left[h_{d,k} + h_{m,k}(\mathbf{\Phi}_b)\right]x_p(t) + n_k(t)$$

where $h_{d,k}$ is the direct propagation response from the router to receiver $k$, $h_{m,k}(\mathbf{\Phi}_b)$ is the metasurface-mediated response under the dual-state coding matrix $\mathbf{\Phi}_b$ ($b \in \{0,1\}$ denoting one of the dual coding states), and $n_k(t)$ is the receiver noise.

To preserve the primary phase-shift-keying (PSK) link, the dual coding matrices $\mathbf{\Phi}_0$ and $\mathbf{\Phi}_1$ are strictly constrained to maintain phase stability, $\Delta\phi = |\angle H_1(\mathbf{\Phi}_1) - \angle H_1(\mathbf{\Phi}_0)| < \varepsilon_\phi$, and sufficient received power, $|H_1(\mathbf{\Phi}_b)|^2 > P_{\text{th}}$. Here, $H_1(\mathbf{\Phi}_b) = h_{d,1} + h_{m,1}(\mathbf{\Phi}_b)$, and $\varepsilon_\phi$ is the allowable phase perturbation determined by the demodulation tolerance of the primary link.

Concurrently, to embed an amplitude-shift-keying (ASK) modulation for the secondary receiver, the states must yield a resolvable amplitude contrast:

$$\Delta A_2 = \left||H_2(\mathbf{\Phi}_1)|^2 - |H_2(\mathbf{\Phi}_0)|^2\right| > \varepsilon_A$$

where $\varepsilon_A$ is the minimum required amplitude contrast. This enables secondary bit recovery via threshold-based envelope detection:

$$\hat{b}(t) = \begin{cases} 1, & |y_2(t)|^2 > \tau \\ 0, & |y_2(t)|^2 \le \tau \end{cases}$$

where $\tau$ is the calibrated decision threshold.

The optimal coding pair is autonomously generated by the digital twin via a constrained search that jointly balances primary-link integrity and secondary-link separability:

$$\{\mathbf{\Phi}_0^*, \mathbf{\Phi}_1^*\} = \arg\max_{\mathbf{\Phi}_0,\mathbf{\Phi}_1} \left[\alpha \min_b |H_1(\mathbf{\Phi}_b)|^2 - \beta\Delta\phi_1 + \gamma\Delta A_2\right]$$

where $\alpha$, $\beta$ and $\gamma$ weight the primary received power, primary phase stability and secondary amplitude contrast, respectively. In our operational configuration, this dual-state switching protocol supported the transmission of a secondary binary image stream at a data rate of approximately 1 Kbps.

**Mobile channel enhancement and physiological monitoring**

For mobile-user enhancement, a ZED 2i binocular camera (providing synchronized RGB images, depth maps, and visual-inertial odometry) tracks the pedestrian's 3D centroid $\mathbf{p}(t) = [x(t), y(t), z(t)]^T$ in the global coordinate frame.

To circumvent online optimization latency, the accessible region is spatially discretized into $N$ nodes $\{\mathbf{r}_i\}_{i=1}^{N}$, each mapped to a precomputed optimal coding matrix $\mathbf{\Phi}_i^*$ via the world model. Real-time trajectory tracking retrieves the nearest-neighbour configuration:

$$i^*(t) = \arg\min_i \| \mathbf{p}(t) - \mathbf{r}_i \|_2, \ \mathbf{\Phi}(t) = \mathbf{\Phi}_{i^*(t)}^*$$

This coordinate-to-code mapping translates continuous movement into low-latency metasurface updates. The resulting performance is quantified by the empirical channel-capacity gain:

$$\Delta C_i = B\log_2\left(\frac{1 + P_i^{\text{on}}/N_i}{1 + P_i^{\text{off}}/N_i}\right)$$

where $B$ is the occupied bandwidth, $N_i$ is the local noise, $P_i^{\text{on}}$ and $P_i^{\text{off}}$ are the received power at the $i$-th spatial node with and without metaEI-WM modulation, respectively. The experimental route, representative coding matrices, predicted field distributions and measured capacity enhancement for the mobile-user trajectory are summarized in Extended Data Fig. 6.

For respiratory monitoring, pedestrian centroid tracking is substituted with anatomical skeletal perception. Using a 34-keypoint human skeleton topology, the system extracts the point located at the chest spine as the dynamic focusing anchor, denoted as $c(t) = [x_c(t), y_c(t), z_c(t)]^T$. A near-field Gerchberg–Saxton algorithm dynamically yields the phase profile maximizing the local field intensity at the chest wall, followed by hardware quantization:

$$\mathbf{\Phi}_{\text{GS}}^*(t) = \arg\max_{\mathbf{\Phi}} |E(c(t); \mathbf{\Phi})|^2$$

The respiration-induced displacement $\delta(t)$ modulates the phase of the backscattered baseband echo $r(t)$, modelled as:

$$r(t) = A_f(t)\exp\left[j\frac{4\pi}{\lambda}\delta(t)\right] + r_{\text{clutter}}(t) + n(t)$$

where $A_f(t)$ is the focused echo amplitude, $r_{\text{clutter}}(t)$ the clutter, and $n(t)$ receiver noise.

Following DC-drift and out-of-band noise suppression, the preprocessed signal $s(t)$ undergoes variational mode decomposition (VMD)[38] into $K$ narrow-band intrinsic mode functions $u_k(t)$ with centre frequency $\omega_k$. Utilizing the alternating direction method of multipliers (ADMM)[39], VMD solves the constrained variational problem:

$$\min_{\{u_k\},\{\omega_k\}} \{\sum_{k=1}^{K} \left\| \partial_t \left[ \left( \delta(t) + \frac{j}{\pi t} \right) * u_k(t) \right] e^{-j\omega_k t} \right\|_2^2 \}, \text{ s.t.} \sum_{k=1}^{K} u_k(t) = s(t)$$

The physiological respiratory mode is subsequently isolated based on spectral energy within the predefined respiration-frequency band $\mathcal{B}_{\text{resp}}$:

$$k^* = \arg \max_{\{k:\, f_k \in \mathcal{B}_{\text{resp}}\}} E_k\,,\ s_{\text{resp}}(t) = u_{k^*}(t)$$

where $f_k = \omega_k/(2\pi)$ and $E_k$ is the energy of the $k$-th mode. Finally, adaptive peak detection within a sliding temporal window filters spurious peaks via amplitude and minimum-interval constraints. The instantaneous respiration rate is evaluated as

$$R_m = \frac{60}{t_p^{m+1} - t_p^{\mathrm{m}}} \text{breaths min}^{-1}$$

where $t_p^{\mathrm{m}}$ and $t_p^{m+1}$ are two adjacent valid peak times. The reported dynamic respiration-rate curve represents the sliding-window median of these instantaneous.

## Data availability

All data needed to evaluate the conclusions in the paper are present in the paper, the Supplementary Materials and the code repository.

## Code availability

The main code used to reproduce the results presented in this paper is available at: https://github.com/ffdzfzh/metaEI-WM-Research-Repository

## Acknowledgements

This work was supported by the National Natural Science Foundation of China (62288101 and 62301146), the National Key Research and Development Program of China (2023YFB3811501), Natural Science Foundation of Jiangsu Province of China (BK20230816), China Postdoctoral Science Foundation (2023M730554 and BX20220065).

## Author contributions

These authors contributed equally: Che Liu and Zhenhao Fu. C.L. and T.J.C. conceived the idea and wrote the paper. Z.F. and C.L. conducted the experiments and prepared the figures. All authors participated in the data analysis and proofreading the manuscript.

## Competing interests

The authors declare no competing interests.

# Figures and Captions

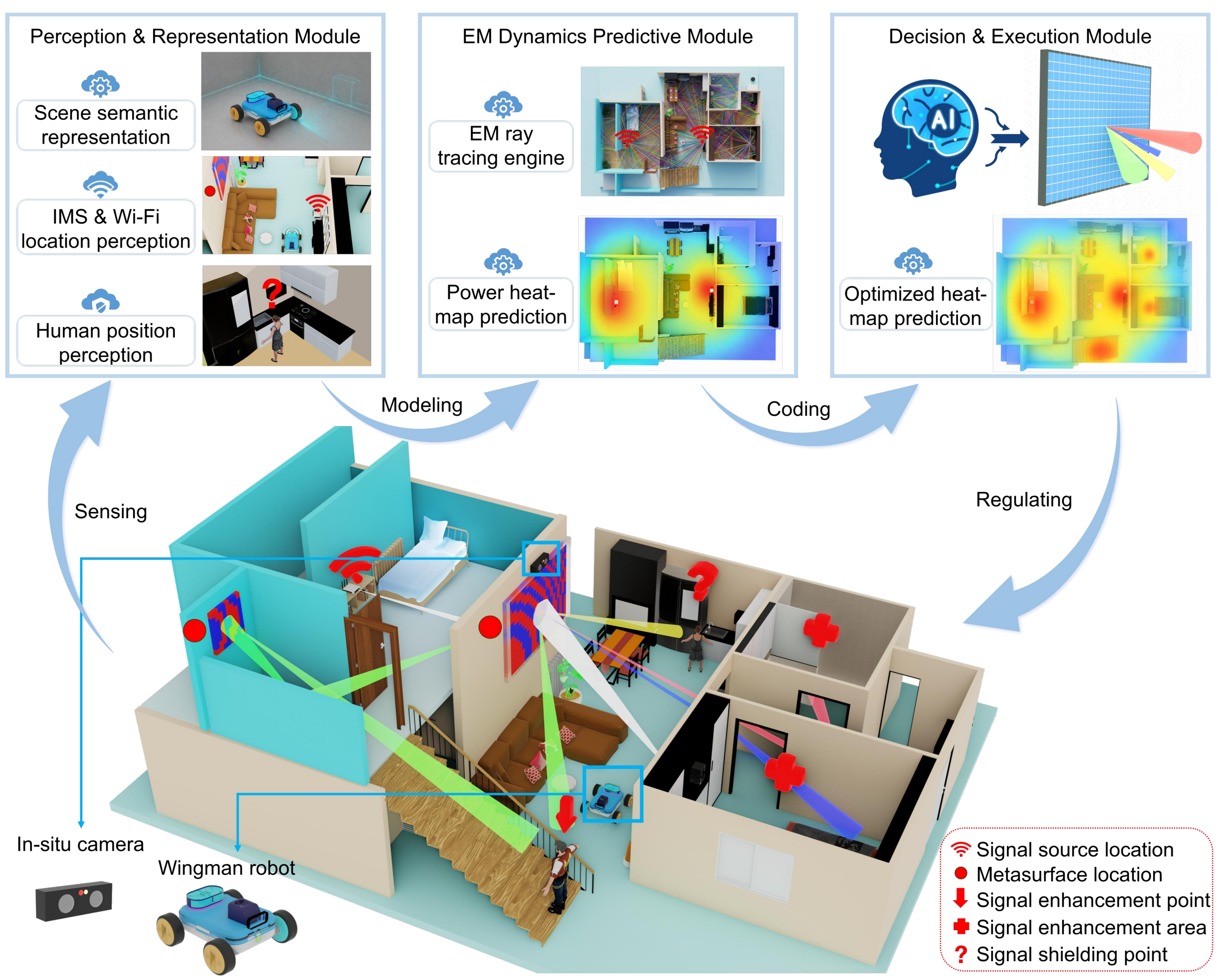

**Fig. 1 | Fundamental architecture of metaEI-WM.** Driven by a vision-language interface, the system integrates three functional modules: a heterogeneous perception and representation module that constructs the spatial semantics for world model representation with a geometry- and material-aware digital twin; an electromagnetic dynamics prediction module that predicts the radio-wave propagation under candidate IMS configurations; and a decision-and-execution module that converts the user-level instructions into hardware-executable IMS configurations.

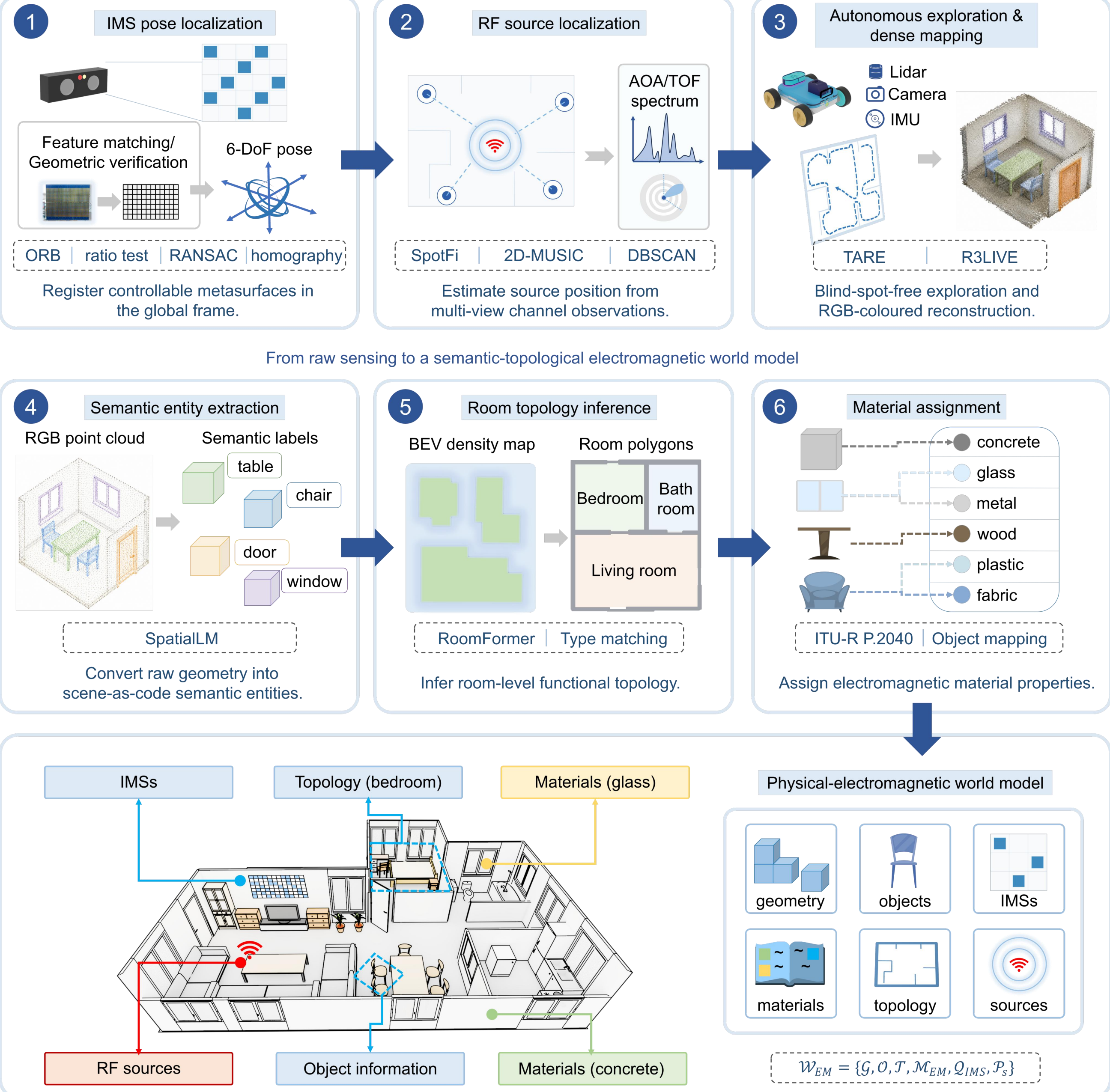

**Fig. 2 | Autonomous construction of the spatial semantics for world model representation.** The six core pipeline stages executed by the cyber-physical agent. (1) IMS pose localization: 6-DoF pose recovery of controllable metasurfaces via ORB features and RANSAC-based homography decomposition. (2) RF source localization: 3D coordinate estimation of radiating entities via 2D-MUSIC and a least-squares ray-intersection estimator. (3) Autonomous exploration and dense mapping: blind-spot-free spatial coverage leveraging the TARE hierarchical framework and R3LIVE LiDAR-camera-IMU fusion. (4) Semantic entity extraction: object-level 3D bounding box and label parsing via SpatialLM. (5) Room topology inference: room-level functional boundary prediction using RoomFormer. (6) Material assignment: automated mapping of semantic labels to frequency-dependent constitutive properties based on the ITU-R P.2040 library. The synthesized physical-electromagnetic world model: the unstructured environment is successfully encapsulated into a searchable, structured representation $\mathcal{W}_{EM} = \{\mathcal{G}, \mathcal{O}, \mathcal{T}, \mathcal{M}_{EM}, \mathcal{Q}_{IMS}, \mathcal{P}_s\}$, pairing digital twins with multi-layer spatial semantics to support subsequent *in silico* electrodynamic forecasting.

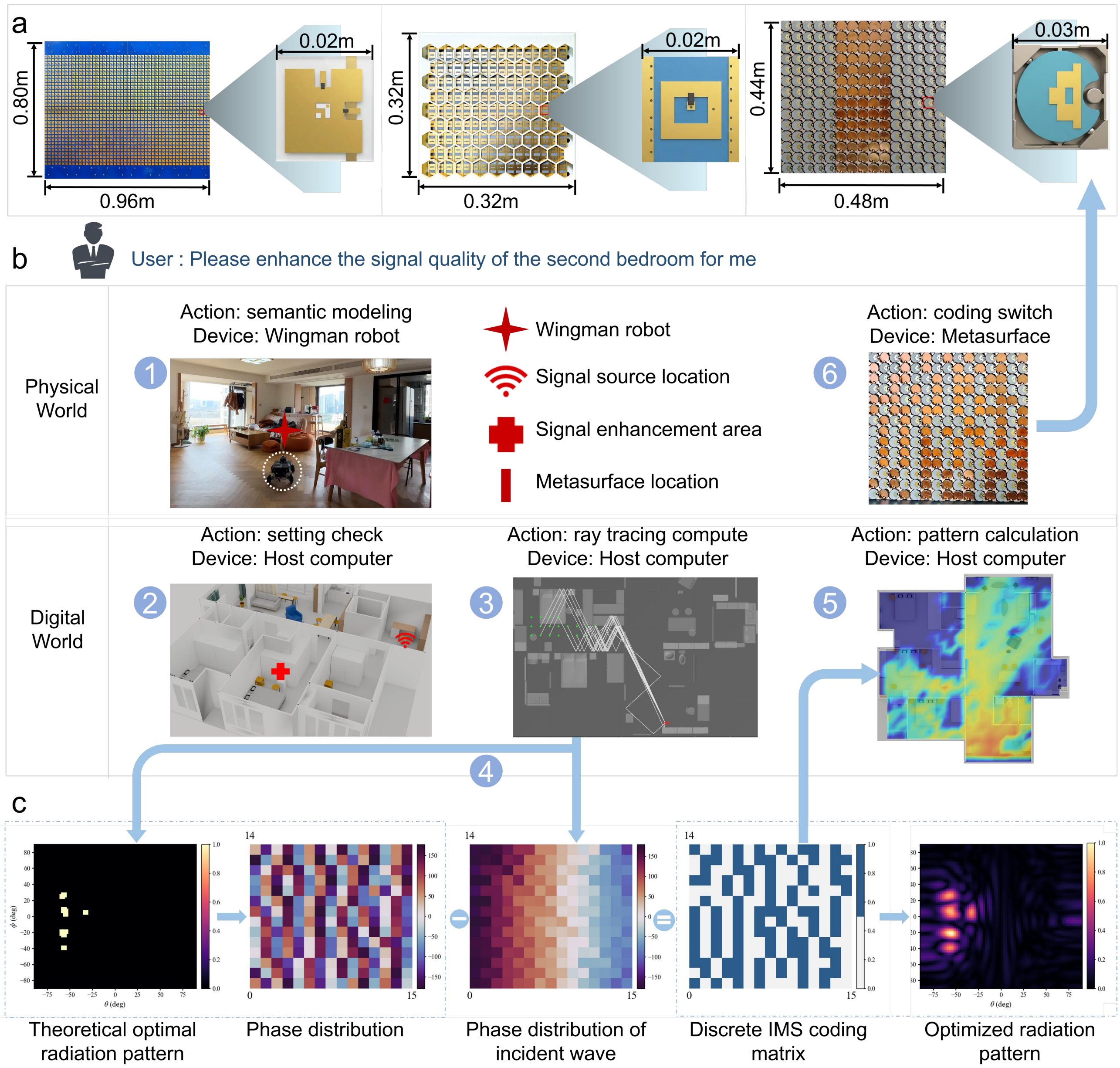

**Fig. 3 | Multimodal operational workflow of the metaEI-WM based on user-specific instructions. a,** Three distinct metasurfaces used for measurement and validation in indoor scenarios. From left to right: 3.5 GHz PIN diode-controlled metasurface (32 rows × 48 columns); 5 GHz PIN diode-controlled metasurface (16 rows × 16 columns); 5 GHz magnetic switch-controlled metasurface (14 rows × 15 columns). **b,** User instruction execution pipeline. The execution workflow triggered by a practical, region-level enhancement request (e.g., "Please enhance the signal quality of the second bedroom for me"). The system leverages the perception and representation module to map the target area, metasurface, and signal transmitter into the word model, proactively resolving spatial ambiguities through human-computer interaction. **c,** Rapid algorithm for IMS coding optimization.

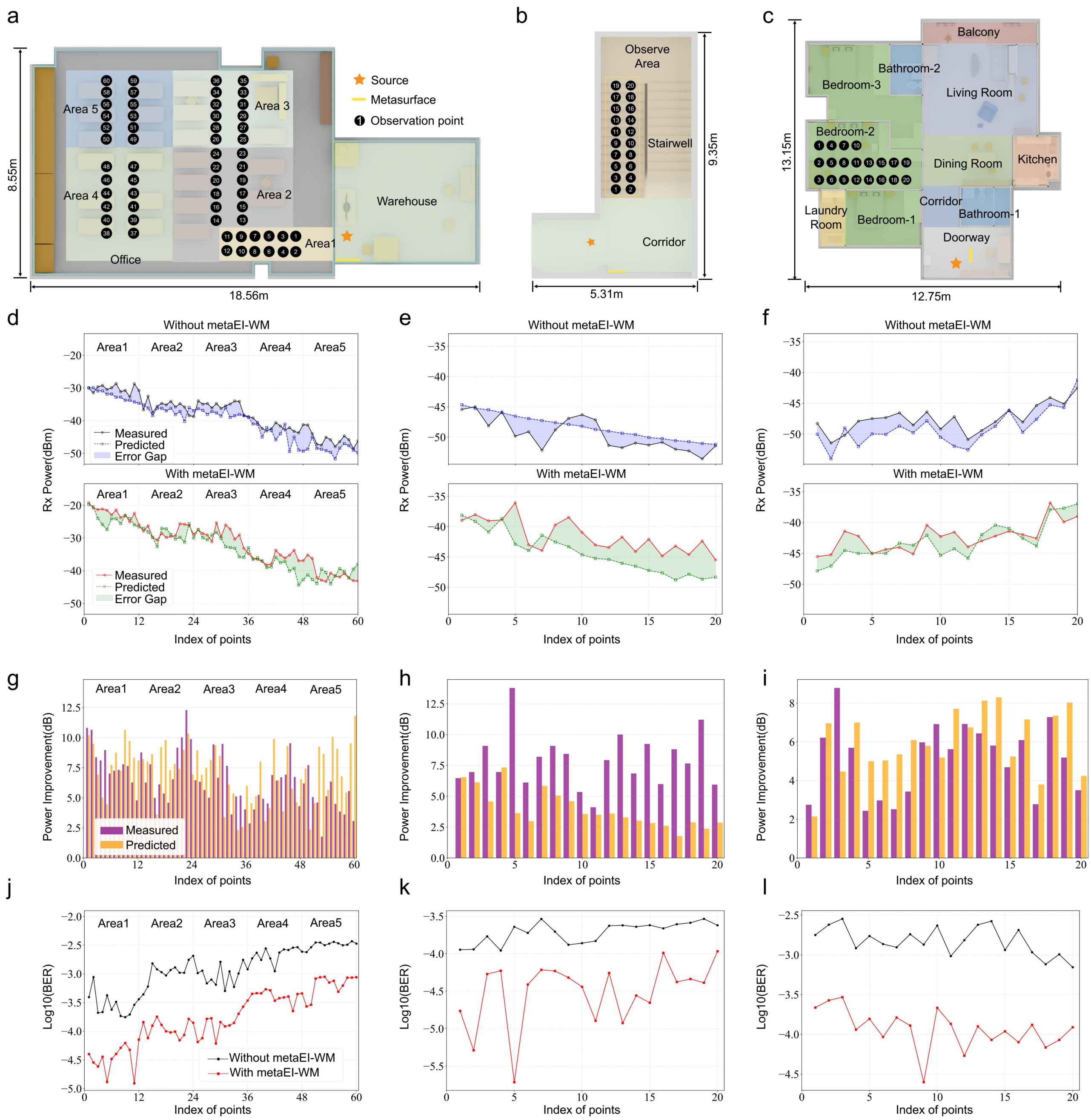

**Fig. 4 | Autonomous dead-zone signal enhancement across the unstructured indoor environments**. **a, b, c,** Semantic-topological schematics of the three testbeds: a workplace with 60 evaluation points across five sub-areas, a multistory corridor with 20 points along a lower staircase , and a residential apartment targeting a deep non-line-of-sight second bedroom with 20 points. **d, e, f,** Validation of predictive fidelity. The distributions of predicted and measured received power before and after IMS modulation exhibit tight agreement between the *in silico* world model and physical reality. **g, h, i,** Simulated versus measured localized power gains. The system autonomously delivered average empirical gains of 6.47 dB, 7.91 dB, and 5.10 dB for the three scenarios, respectively. **j, k, l,** Link-level reliability improvements. Following world-model-guided optimization, the physical links achieved bit error rate (BER) reductions of 0.84, 0.80, and 1.11 orders of magnitude across the respective environments.

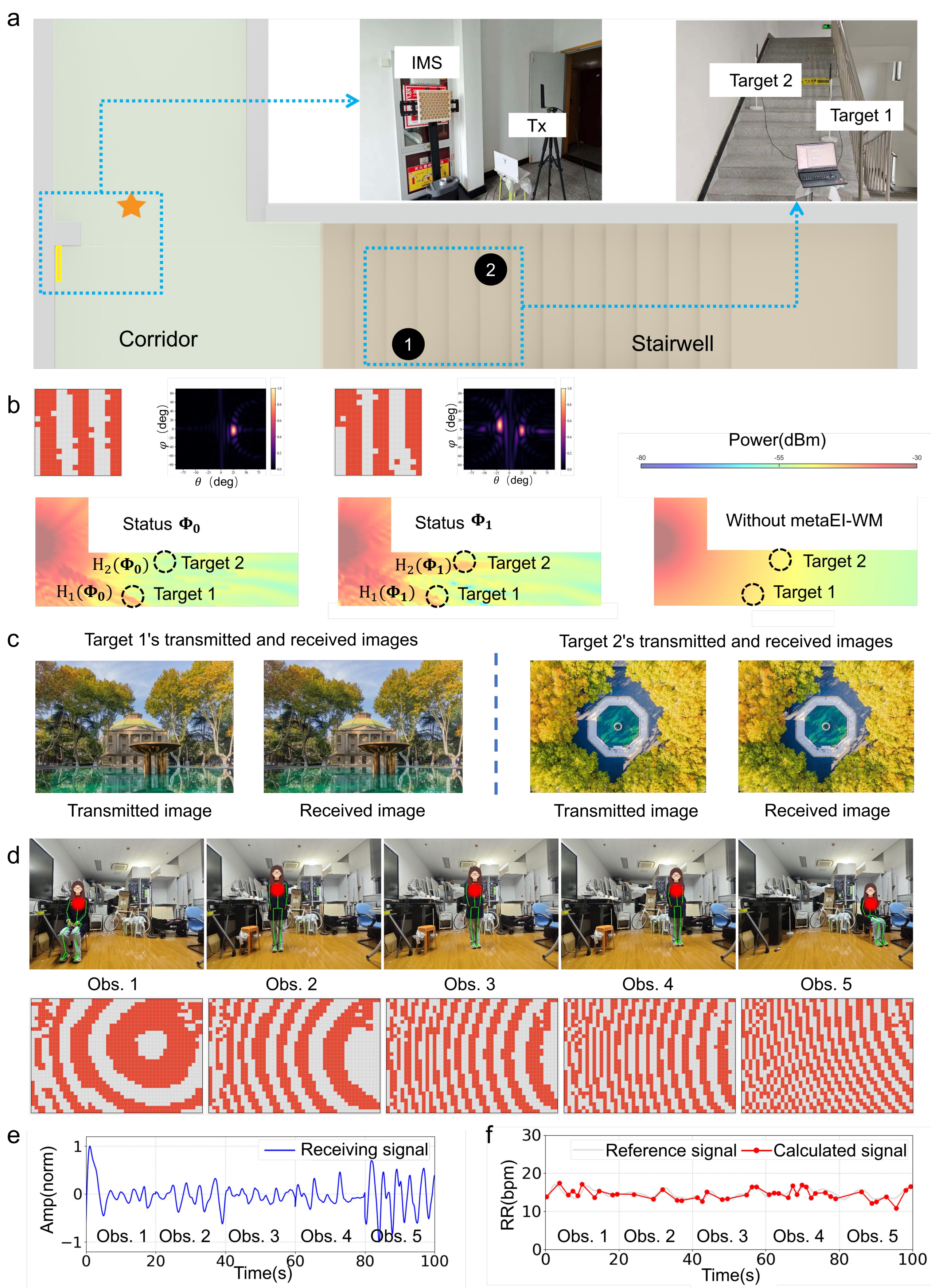

**Fig. 5 | Symbiotic wireless communication and mobile physiological monitoring enabled by metaEI-WM. a,**

Experimental configuration for symbiotic communication in a corridor–staircase environment, where the router, IMS and two receivers form a shared electromagnetic ecosystem. Target 1 receives the IMS-enhanced primary Wi-Fi link, whereas Target 2 receives the IMS-scattered secondary signal. **b,** Dual-state IMS coding strategy generated by metaEI-WM and its predicted spatial field distributions . The two coding states preserve the focused field and phase stability at Target 1 while producing distinguishable field amplitudes at Target 2, enabling amplitude-shift-keying modulation of the secondary backscattered stream. **c,** Image-transmission results, showing coexistence of the protected primary link and the metasurface-mediated secondary link. **d,** Schematic of beam focusing and tracking results for respiratory monitoring. As the user moves, the system dynamically adjusts the metasurface coding to focus electromagnetic waves on the user's chest in real time. **e,** Time-domain wireless echoes collected during mobile respiratory monitoring as the user moves from Observation Point 1 to Observation Point 5. **f,** Dynamic respiration rate estimated from the metaEI-WM-optimized wireless signal, demonstrating close agreement with a wearable reference device.

# Extended Data Figures and Table

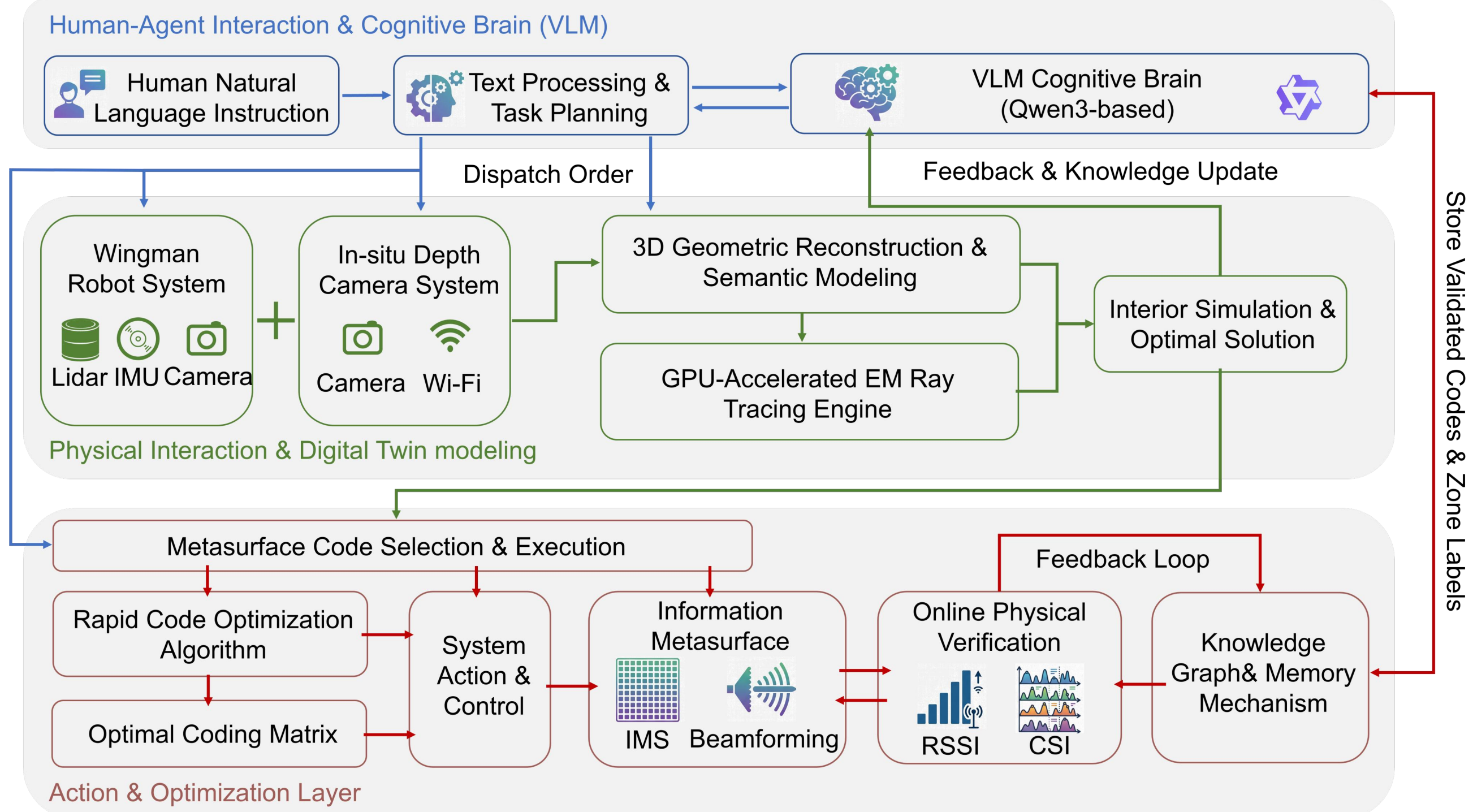

**Extended Data Fig. 1 | The operational pipeline of the metaEI-WM system.** The VLM cognitive brain decomposes human natural language instructions into executable sub-tasks. Heterogeneous sensing network collects multimodal data to build spatial-semantic world model representation with a geometry- and material-aware digital twin. A GPU-accelerated EM ray-tracing engine is adopted to predict radio propagation maps. Combined with optimization algorithms, it yields the theoretically optimal coding strategy for IMS development.

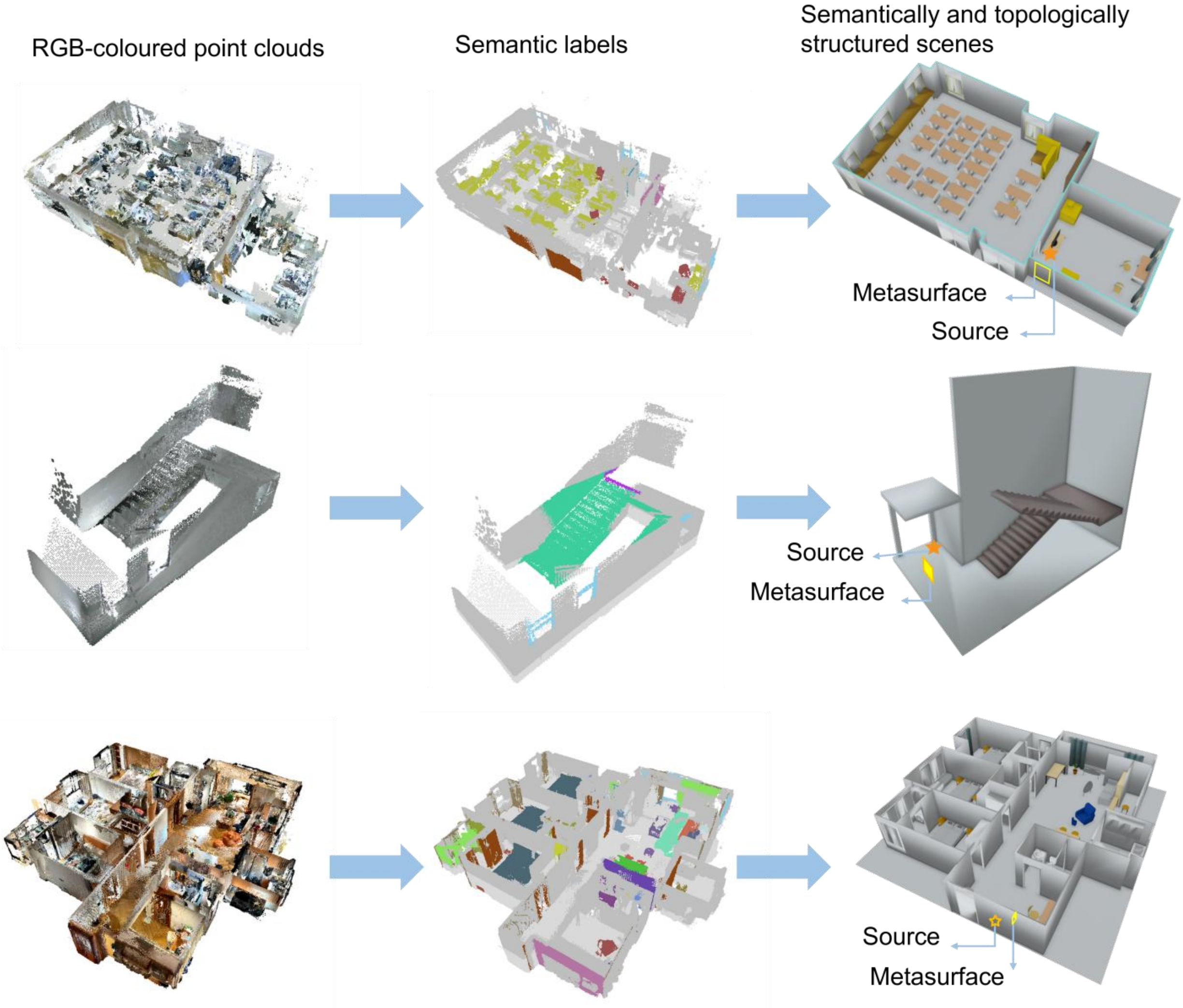

**Extended Data Fig. 2 | The semantically and topologically structured scene of indoor environments.** By integrating electromagnetic material properties into the structured scene, it evolves into a spatial-semantic world model representation encompassing a geometry- and material-aware digital twin.

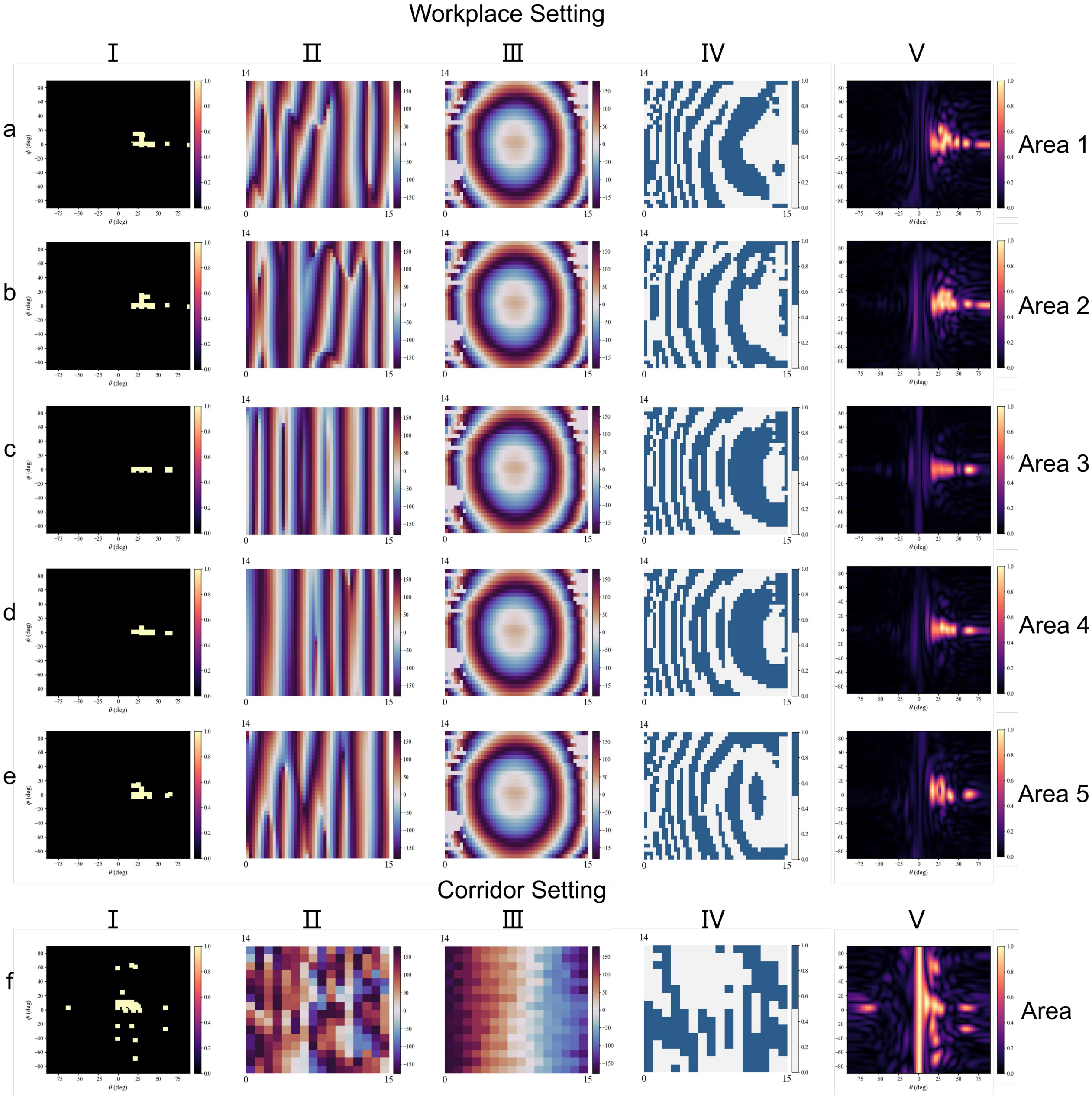

**Extended Data Fig. 3 | Coding optimization procedures and outcomes in the workplace and corridor.** Roman numerals **I** to **V** denote the following elements for each scenario: **I**. Theoretical optimal radiation pattern at the IMS aperture; **II**. Phase distribution at the IMS aperture; **III**. Phase distribution of incident wave; **IV**. Discrete IMS coding matrix; **V**. Optimized radiation pattern at the IMS aperture. Letters **a** to **f** indicate the optimization processes for seven subregions within the two main scenarios. The optimization process for the residential setting is presented in Fig. 2b.

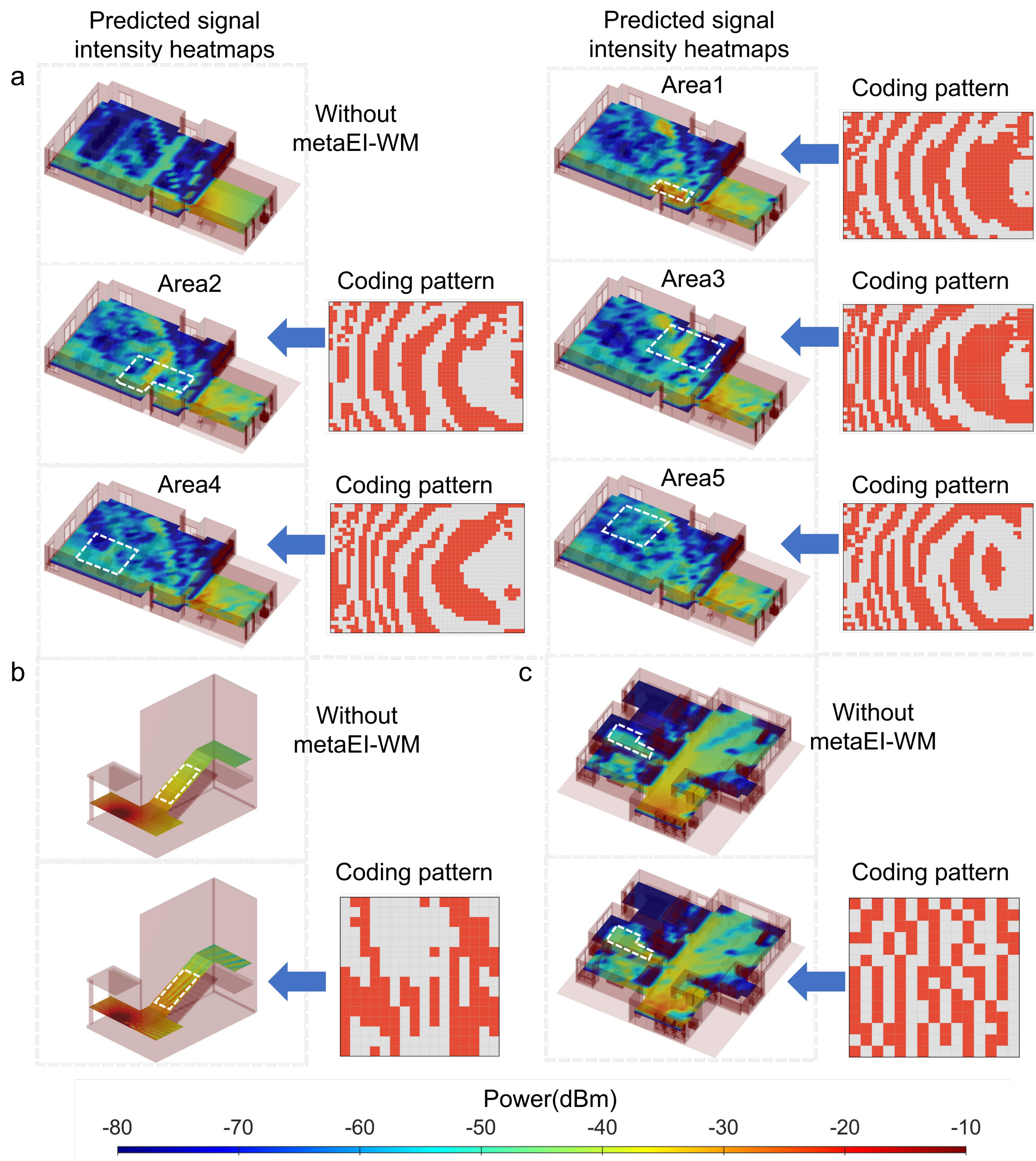

**Extended Data Fig. 4 | Predicted radio coverage across three representative indoor environments. a,** Area-specific metasurface coding configurations and corresponding predicted three-dimensional radio-coverage heatmaps in the office environment, showing the unmodulated baseline and targeted signal enhancement across five subareas. **b,** Predicted signal-intensity heatmaps and corresponding metasurface coding configurations in the corridor and staircase environment before and after language-prompted metaEI-WM intervention. c, Predicted signal-strength distributions across the residential apartment before and after metaEI-WM modulation, with the second bedroom highlighted by a white dashed rectangle.

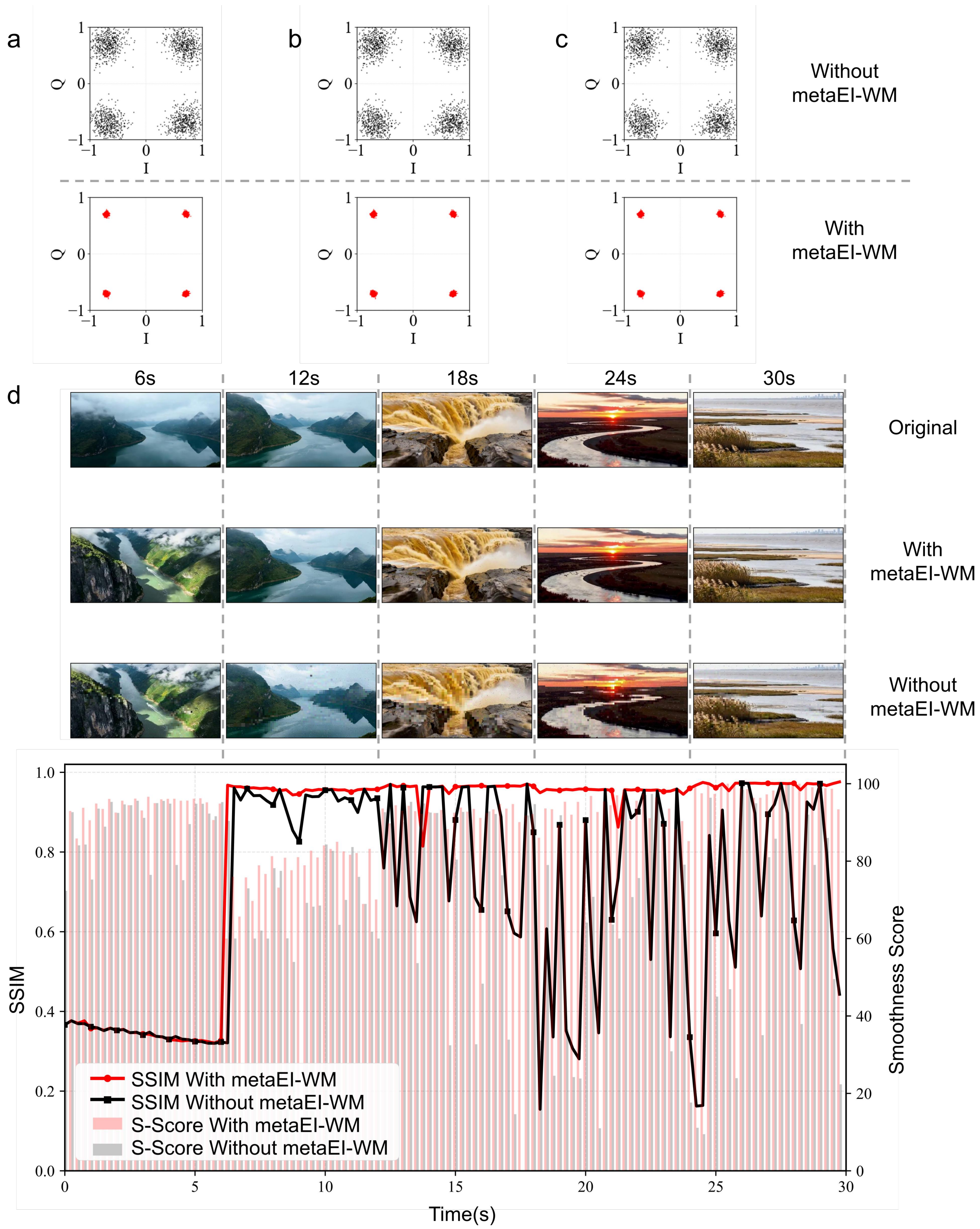

**Extended Data Fig. 5 | Modulation-domain constellation recovery and application-level video-streaming quality. a–c**, Received I/Q constellation diagrams before (top) and after (bottom) world-model-guided optimization at representative evaluation points within the workplace (**a**, 24th point), multistory corridor (**b**, 20th point), and residential apartment (**c**, 5th point). The transition from highly dispersed, multipath-distorted distributions to compact, linear clusters demonstrates consistent link-level phase and amplitude recovery across all testbeds. **d**, Real-time video-streaming performance over a 30-s transport stream in the residential apartment link. The

structural similarity index measure (SSIM, left axis) and empirical smoothness score (S-Score, right axis) quantify the playback quality with (red line/pink bars) and without (black line/grey bars) metaEI-WM modulation. Higher S-Scores denote the successful suppression of temporal freezing and mosaic artifacts, as visually corroborated by the representative video frames (middle panels) captured at 6-s intervals.

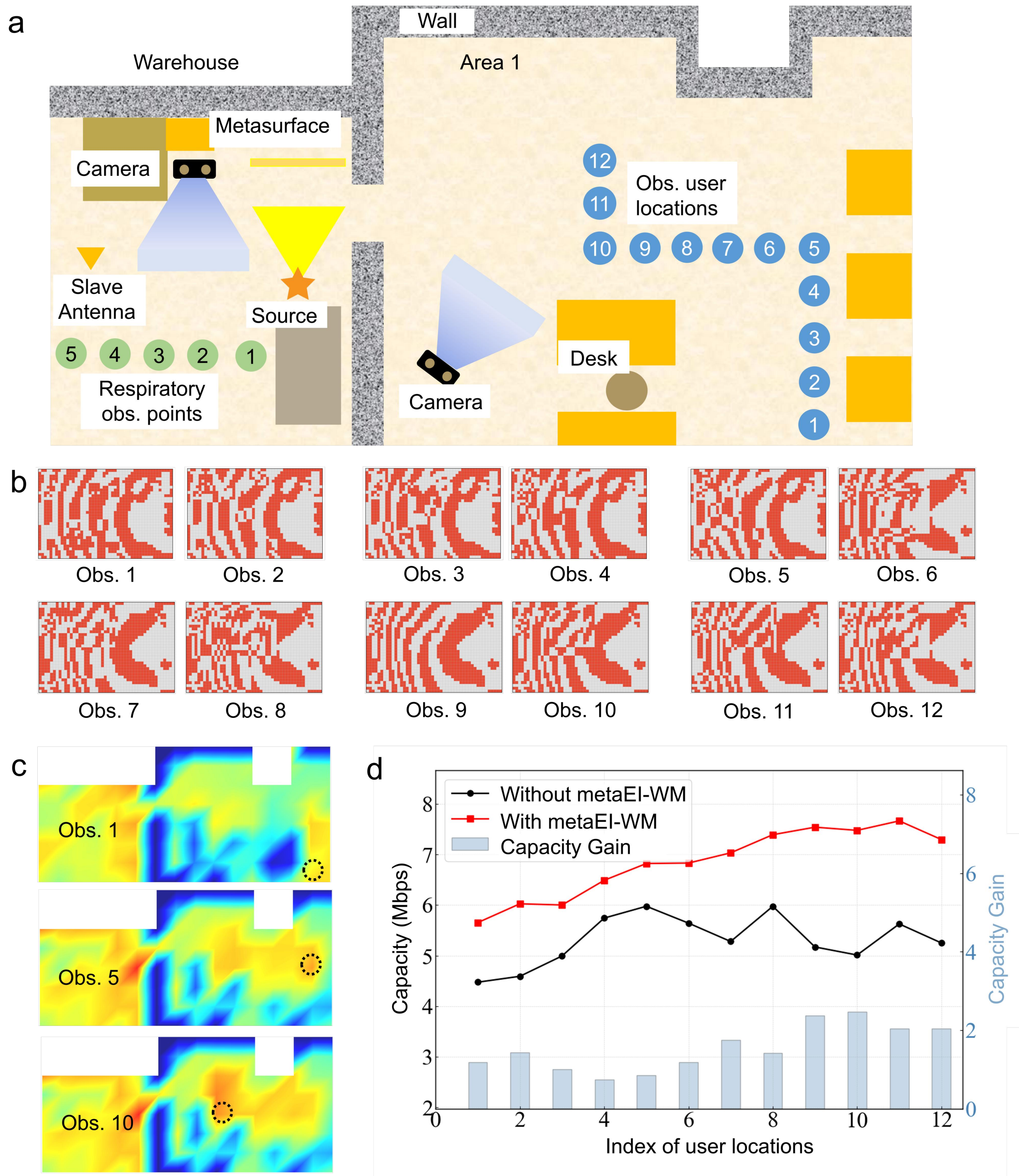

**Extended Data Fig. 6 | Validation of dynamic signal enhancement for mobile users using metaEI-WM. a,** Experimental configuration for signal amplification in Area 1 and respiration-monitoring measurements in warehouse. Numbered circles indicate the spatial trajectory of the mobile user. **b,** System-derived optimal metasurface coding patterns for 12 representative user locations in Area 1. **c,** Predicted signal-intensity heatmaps for observed user locations 1, 5 and 10. **d,** Experimentally measured channel-capacity enhancement at the 12 user locations in Area 1.

**Extended Data Table 1 | Scenario-level and sub-area-level statistics for the metaEI-WM-enabled dead-zone recovery.**

| Scenario | Office | Corridor–staircase | Residential apartment |
|---|---|---|---|
| Number of sampling points (N) | 60 | 20 | 20 |
| Simulated Rx power without metaEI-WM (dBm) | -40.08 ± 6.15 | -48.32 ± 2.05 | -49.04 ± 2.90 |
| Simulated Rx power with metaEI-WM (dBm) | **-32.91 ± 6.66** | **-44.36 ± 3.42** | **-43.05 ± 3.02** |
| Measured Rx power without metaEI-WM (dBm) | -37.90 ± 5.69 | -49.51 ± 2.64 | -47.49 ± 2.23 |
| Measured Rx power with metaEI-WM (dBm) | **-31.43 ± 6.91** | **-41.60 ± 2.67** | **-42.39 ± 2.26** |
| Simulated power gain (dB) | 7.17 | 3.95 | 5.99 |
| Measured power gain (dB) | 6.47 | 7.91 | 5.10 |
| BER without metaEI-WM | $1.615\times10^{-3}$ | $2.018\times10^{-4}$ | $1.616\times10^{-3}$ |
| BER with metaEI-WM | **$2.853\times10^{-4}$** | **$4.250\times10^{-5}$** | **$1.335\times10^{-4}$** |
| BER reduction | 0.84 orders | 0.80 orders | 1.11 orders |
| IOA without metaEI-WM | 0.9316 | 0.9205 | 0.8312 |
| IOA with metaEI-WM | **0.9405** | **0.8366** | **0.8655** |

Supplementary Information for

# Metasurface embodied intelligence through electromagnetic world model

Che Liu[1,†,*], Zhenhao Fu[1,†], Qian Ma[1], Jiajing Wu[1], Wen Ming Yu[1], Lianlin Li[2] and Tie Jun Cui[1,3,4,*]

[1]State Key Laboratory of Millimeter Waves, Southeast University, Nanjing 211189, China

[2]State Key Laboratory of Photonics and Communications, Peking University, Beijing 100871, China

[3]Institute of Electromagnetic Space, Southeast University, Nanjing 211189, China

[4]Center of Metamaterials, Suzhou National Laboratory, Suzhou 215164, China

*E-Mail: cheliu@seu.edu.cn, tjcui@seu.edu.cn

[†]Equally contributed to this work.

**Outline**

## Supplementary Note 1. The Implementation of metaEI-WM Cognitive Brain

To ensure the efficient, stable, and secure operation of highly complex embodied intelligence systems in physical environments characterized by the collaboration of multiple heterogeneous hardware components, this supplementary note systematically elaborates on the specific implementation scheme of the metaEI-WM Cognitive Brain. The system employs a vision-language cognitive hub deployed on edge computing nodes as the global decision-making core. Simultaneously, to guarantee the reliable operation of this hub, the system introduces a multi-process virtual environment isolation mechanism at the bottom layer to address software and hardware dependency conflicts, while deploying a privacy protection strategy based on cloud-edge decoupling within the top-layer interaction.

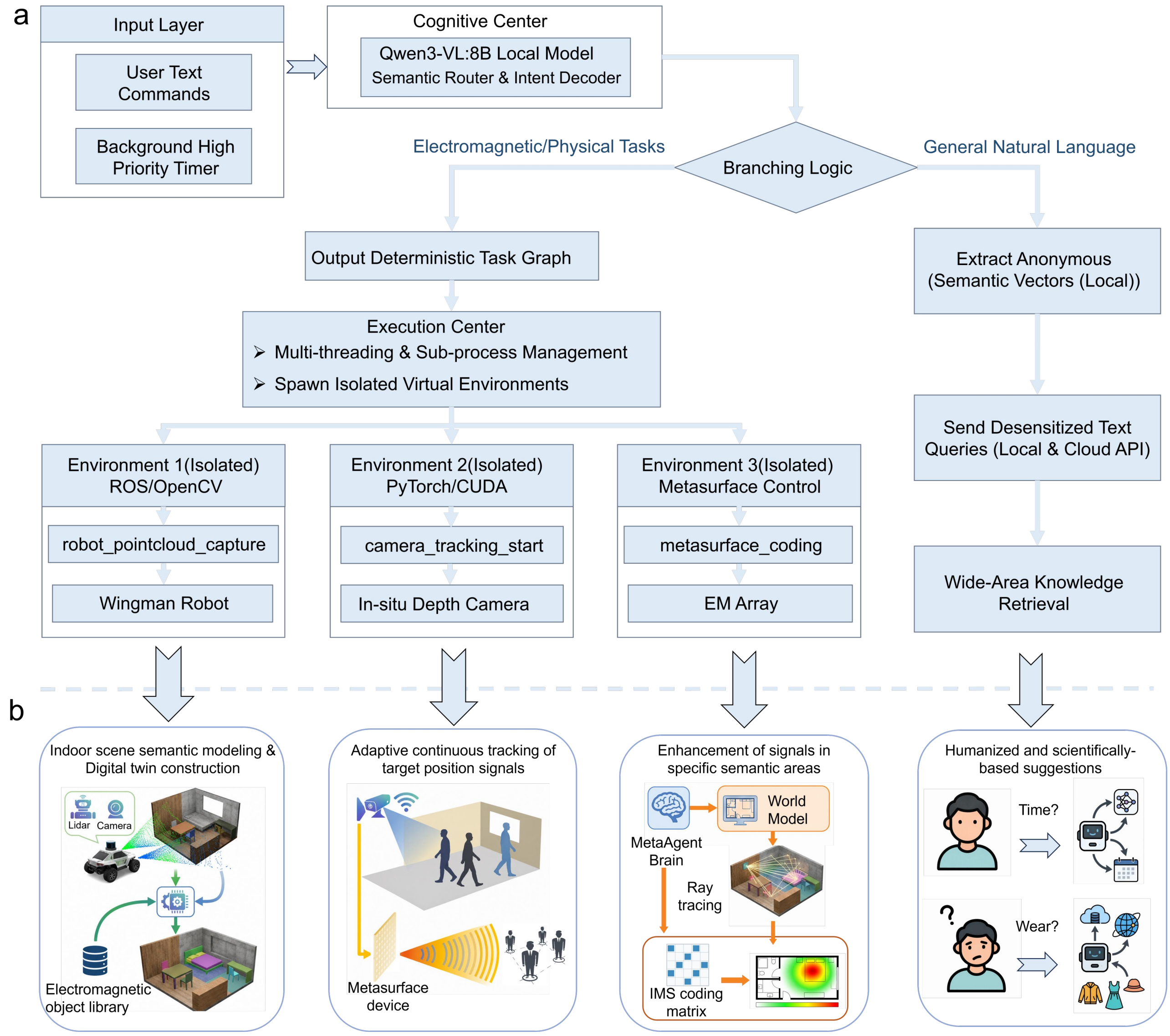

**Fig S1. The metaEI-WM cognitive center & isolated execution framework architecture diagram. (a)** This figure illustrates the complete data flow from the input layer to the final task execution within the system. The core edge node employs the Qwen3-VL:8B local model as a semantic router. After classifying intents, the execution center securely schedules underlying hardware (Wingman robot, depth camera, and electromagnetic array) within isolated virtual environments, or handles general natural language tasks accordingly. **(b)** The four highly autonomous operation modes: indoor scene semantic modeling, adaptive continuous tracking, specific area signal enhancement, and general natural language tasks.

## 1. Cognitive Center and Cross-Modal Semantic Abstraction Based on Edge-Localized Qwen3-VL

In the system architecture of metaEI-WM (as illustrated in Fig. S1), the selection and deployment location of the cognitive center are directly related to the system's response latency, spatial perception accuracy, and the safety of physical execution. This system abandons the traditional cloud

API invocation approach and chooses to deploy the Qwen3-VL:8B vision-language foundation model directly at the local edge computing node. This architectural decision is primarily based on considerations of the stringent physical constraints of 6G cyber-physical systems:

- Elimination of transmission latency: In high-frequency physical communication scenarios requiring intervention in electromagnetic wave propagation paths, the system is highly sensitive to the transmission time of control commands. Localized deployment thoroughly eliminates the network transmission latency and uncertainty jitter of cloud round-trips at the physical level, enabling the agent to perform real-time radio frequency (RF) channel reconstruction and metasurface beam tracking for spatially moving targets within a millisecond-level tolerance.
- Guarantee of data security: Localized processing prevents the external transmission of indoor depth images and spatial point cloud data collected by the system, ensuring that the agent operates compliantly in highly confidential research institutes or private domestic environments with strict privacy requirements.
- Deterministic task graph decoding: The cognitive center deeply integrates the capabilities of Qwen3-VL:8B in cross-modal data semantic abstraction and long-horizon task decomposition. During the deterministic task graph decoding phase, the model is strictly confined as a structured semantic router and intention decoder, stripped of open-ended free generation privileges. Through context matching, the semantic router maps the user's unstructured macroscopic commands (e.g., "Help me enhance the WiFi signal in the bedroom") into a predefined and physically verified atomic operation ontology within the system. The intention decoder is responsible for extracting key entities (such as target location, user identity, and desired signal gain type) and generating corresponding structured state parameters, thereby safely projecting the open semantic space into a closed system execution state space.

Through the aforementioned neuro-symbolic routing mechanism, long-horizon unstructured commands are autonomously decomposed into a "directed acyclic task graph" spanning physical perception, visual computation, and electromagnetic manipulation (see Fig. S1a). It is worth noting that the abstract subtasks generated by the large model exist solely at the logical decision-making layer. The system does not rely on the model to generate underlying hardware control scripts in real-time; instead, task graph nodes precisely trigger pre-hardcoded and sandbox-tested Python control scripts and API handles. This design effectively circumvents the uncertainties associated with direct code generation by large models, ensuring underlying physical safety.

Although the cognitive hub provides robust capabilities for high-level intent decomposition, translating these intents safely and stably into bottom-layer physical actions and adapting them to broad-area home interaction scenarios necessitates the support of specific bottom-layer execution frameworks and top-layer data mechanisms.

**2. Multi-Threaded Environment Isolated Execution Framework**

To ensure the stability of the cognitive hub when cooperatively scheduling multiple heterogeneous devices within the physical space, a fine-grained environment isolation framework is designed at the bottom layer of metaEI-WM. Specifically, the low-level scripts that drive the wingman robot for indoor roaming and LiDAR point cloud acquisition are heavily dependent on a specific, older version of the Robot Operating System (ROS) and its associated older version of the OpenCV computer vision library. Concurrently, the in-situ depth camera adaptive tracking module, responsible for the real-time capture of dynamic targets, along with the subsequent large model inference module for extracting three-dimensional spatial features, strictly require the latest version of the PyTorch deep learning framework, highly parallel GPU computation graphs, and matching low-level CUDA acceleration driver libraries. Furthermore, the FPGA control scripts utilized for controlling the phase of metasurface array units demand exclusive access to specific serial communication protocols and timing logic drivers.

If all these cross-modal control, sensing, and AI inference codes were forcefully mounted or imported into a single Python global runtime environment during system architecture design, it would inevitably lead to severe Dynamic Link Library (DLL) version conflicts, environment variable pollution, and broken package dependency trees (Dependency Hell). For instance, the Python bindings of older ROS versions are highly susceptible to symbol conflicts with the underlying C++ dependencies of the latest PyTorch versions, leading to core dumps during runtime and causing the entire agent system to paralyze.

To thoroughly resolve this heterogeneous integration challenge, the execution center of metaEI-WM discards the traditional monolithic code integration approach and creatively leverages the multi-threaded asynchronous non-blocking architecture and highly refined child process management mechanisms of modern operating systems. The system maintains the overall state machine via a global master daemon thread, allocating completely isolated virtual runtime environments for each specific heterogeneous hardware invocation.

In specific engineering implementations, the master daemon thread instantly generates a runtime environment with an independent namespace for specific underlying sensing or manipulation scripts upon awakening, achieved by precisely capturing and mapping the absolute paths of different Python interpreters or dynamically calling the internal sys.executable attribute of the system. For example:

- Isolated Environment 1 (Robot Navigation and Sensing): The master process initiates a child process mounted with ROS and legacy computer vision libraries, exclusively responsible for executing mobile acquisition tasks such as robot_pointcloud_capture(), with its internal data flow completely independent of the deep learning environment.
- Isolated Environment 2 (Vision and Spatial Inference): The master process spins up a PyTorch environment calling the latest CUDA acceleration libraries, dedicated to executing camera_tracking_start(), ensuring that the visual stream inference of the depth camera holds the highest frame rate and computational priority.
- Isolated Environment 3 (Real-Time Electromagnetic Component Manipulation): An ultra-low latency, lightweight I/O isolated environment responsible for receiving phase matrices via high-speed buses and executing metasurface_coding(), interfacing directly with controllers like FPGAs.

To address the issue of environment variable transmission caused by cross-environment scheduling, the master scheduler explicitly rewrites and injects PATH and PYTHONPATH to ensure each child process obtains exclusive dependency paths. Data interaction between different isolated environments (e.g., the vision module outputting 3D coordinates to the electromagnetic control module) strictly relies on operating system-level Inter-Process Communication (IPC) mechanisms or memory-safe queues. This multi-tiered isolation architecture significantly enhances system robustness; even if a single child process crashes due to a sensor disconnection or Out-Of-Memory (OOM) error, the master state machine remains undisturbed and can capture the anomaly, attempt a restart, or switch to backup perception modalities within milliseconds.

**3. Compatibility and Security Decoupling of General Interaction and Smart Home Ecosystems**

While resolving bottom-layer physical execution conflicts, to enable the cognitive hub to operate securely and compliantly within a broader smart home ecosystem, the system implements an exceedingly stringent security decoupling mechanism at the top-layer data interaction level. Although metaEI-WM demonstrates rigorous determinism and professionalism in structured physical manipulation tasks, as a future communication core component dedicated to seamless integration into smart home ecosystems and complex industrial architectures, it inevitably needs to handle a vast amount of open-ended natural language interactions unrelated to immediate electromagnetic manipulation. When the internal semantic router determines that the received user text or voice command (such as asking for weather, medical advice, or daily chatting) falls outside the scope of RF physical operations, the metaEI-WM automatically switches into a non-physical execution state —a general chatbot mode known as the "Family Care" mode. In this mode, the system analyzes the context through retrieval-augmented mechanisms and presents highly humanized and scientifically grounded feedback.

To maximize the system's generalized intelligence while completely eradicating the risk of privacy leaks, metaEI-WM strictly enforces a "cloud-edge decoupling" data security strategy and semantic anonymization mechanism under the "Family Care" mode. In this architecture, a logical firewall is set between edge nodes and cloud services, and all sensitive multimodal raw data is confined to processing within local hardware. The local cognitive center acts as an "information purifier," extracting core intentions from visual and acoustic data and applying dimensionality reduction obfuscation through mechanisms like differential privacy. This transforms commands into anonymous texts devoid of specific identities or spatial attributes.

Only purely text-based requests that have undergone multi-level verification and complete desensitization are permitted to cross the firewall to invoke cloud-based knowledge retrieval APIs. Once the cloud generates and returns the answer, the local VLM secondarily integrates it with current environmental data to generate the final response presented to the user. This collaborative architecture ensures compliance with “data staying on-device” while effectively preserving the large model’s general retrieval capabilities in the wide-area knowledge domain.

**4. Semantic Decomposition and Mapping Logic of Typical Physical Tasks**

Having established the vision-language cognitive hub as the global brain, along with the bottom-layer isolation framework and top-layer security decoupling strategies that support its operation, the metaEI-WM system forms a comprehensive multi-agent collaborative state machine mechanism (summarized as the four highly autonomous operation modes in Fig. S1b). The following section provides an in-depth tabular and narrative deconstruction of the system's semantic instruction output, bottom-layer execution actions, and the isolation mapping logic of the virtual environment, contextualized within three typical high-level task scenarios:

**Task 1:** *Geometric Topology Extraction and Construction of a geometry- and material-aware digital twin*

Constructing a geometry- and material-aware digital twin of the indoor physical environment is the cornerstone for realizing subsequent electromagnetic manipulation. This task requires the agent to drive a highly maneuverable wingman robot to patrol the indoor space, utilize onboard LiDAR or depth cameras to capture point cloud data, and ultimately abstract a 2D topology map suitable for radio planning. This composite task is also included as a prerequisite in the "regional signal enhancement" task in the main text, with its core associated invocation module being scene_reconstruction(). The specific semantic command outputs, underlying execution actions, and associated calling modules for this process are detailed in Table S1.

**Table S1** Semantic decomposition and execution mapping for construction of a geometry- and material-aware scene model of the physical space

| **Agent Semantic Com-mand Output** | **Underlying Execution Action & Environ-mental Isolation Strategy** | **Associated Calling Mo-dule** |
|---|---|---|
| *"Detect the current status of the wingman robot"* | Hardware status self-check: Detect the robot's underlying connection status and hardware handshake signal integrity | robot_check() |
| *"Wingman robot launch-ed, work is about to begin"* | Communication link initialization: Establish an SSH remote control channel with the robot's computing center via encrypted protocols | robot_launch() |
| *"Wingman robot is ac-quiring scene point clouds"* | Environmental sub-process scheduling: Trigger the robot's underlying sensor drivers and initiate high-bandwidth data stream capture | robot_pointcloud_capture() |
| *"Inferring and visuali-zing the acquired point cloud scene"* | Cross-environment sub-process scheduling: Invoke an independent deep learning framework environment for spatial geometric and semantic inference | pointcloud_inference() |
| *"Applying the electro-magnetic object library for matching and reconstruction"* | Physical property mapping: Extract semantic labels from structured parsing results, match preset electromagnetic physical libraries, and render materials | EM_objects_match() |
| *"Acquiring the 2D scene floor plan and executing topological area divi-sion"* | Geometric topology extraction: Parse multi-room boundaries and spatial connectivity from polygon sets to reconstruct the 2D topological structure | Plane_obtain() |
| *"Indoor scene modeling complete"* | State machine reset: Clear the task queue, transition the cognitive center state, and switch the system to idle listening mode | Cognitive center state transition |

**Task 2:** *Depth Camera Perception of Dynamic Moving Targets and Adaptive Tracking of Metasurface Signals*

Unlike static scene scanning that relies on mobile robots, for the task of "please complete the signal enhancement in the area where the person is located," the core of the system's perception network immediately and smoothly transitions to a fixed-deployment, in-situ, high-frame-rate depth camera network. In this task, metaEI-WM faces extremely high temporal coupling requirements: the agent must map the spatial displacement parameters of the target human body captured by the visual system into electromagnetic phase deflection control commands for every patch element of the metasurface array in real-time, with near-zero latency. The execution logic and environmental isolation strategy for adaptive tracking are summarized in Table S2.

**Table S2** Semantic decomposition and execution mapping for dynamic target perception and adaptive signal tracking

| Agent Semantic Com-mand Output | Underlying Execution Action & Environ-mental Isolation Strategy | Associated Calling Mo-dule |
|---|---|---|
| *"Detect the current working status of the depth camera"* | Hardware status self-check: Detect the physical connection status and operational availability of the in-situ depth camera | tof_camera_check() |
| *"Depth camera launch-ed, work is about to begin"* | Sensor initialization: Load camera driver parameters, initiate device connection and data bus initialization | tof_camera_launch() |
| *"Depth camera is work-ing"* | Cross-environment sub-process scheduling: Isolate and activate the deep learning-based computer vision real-time inference pipeline | camera_tracking_start() |
| *"Switch metasurface co-ding to complete contin-uous adaptive tracking of the target"* | Cross-environment sub-process scheduling: Real-time calculation and rewriting of the metasurface array control words based on the spatial coordinates output by the vision module | metasurface_adaptive_coding() |

This task perfectly demonstrates the superiority of the isolation architecture. Heavy visual neural network inference is locked within the GPU-accelerated environment, while the lightweight yet highly latency-sensitive phase matrix dispatch operations are isolated in an independent I/O environment. The two exchange coordinate data via high-speed memory pipes, ensuring coordinated action between visual perception and electromagnetic regulation.

**Task 3:** *Autonomous Micro-Doppler Respiratory Detection Mode Under Health Guardian Mechanism*

The design vision of metaEI-WM extends beyond being a responsive command execution tool; it aims to build an intelligent wireless ecosystem with proactive perception capabilities. The trigger mechanism for the Task 3 "Health Guardian" mode is particularly unique: it is not awakened by explicit user voice commands or text inputs, but rather proactively initiated via an interrupt request thrown by a background high-priority time cycler embedded in the system's underlying state machine. In this mode, the system combines computer vision and micro-Doppler radar signal processing technologies, utilizing the metasurface to proactively focus RF energy to achieve non-contact, imperceptible monitoring of faint physiological signs of the human body. The detailed task breakdown for this proactive monitoring mode is presented in Table S3.

**Table S3** Semantic decomposition and execution mapping for autonomous micro-doppler respiratory detection

| Agent Semantic Com-mand Output | Underlying Execution Action & Environ-mental Isolation Strategy | Associated Calling Mo-dule |
|---|---|---|
| *"metaEI-WM entering 'Health Guardian' mode"* | Environment takeover and initialization: Automatically detects the current depth camera state and forcefully initiates hardware connection | tof_camera_check(), tof_camera_launch() |
| *"Performing human skeletal keypoint extraction"* | Cross-environment child process scheduling: Loads the posture model and continuously executes human skeletal keypoint extraction | bone_node_extraction() |
| *"Switching metasurface coding to achieve adaptive continuous tracking of the target"* | Cross-environment child process scheduling: Calculates and rewrites the state of metasurface array units in real-time based on spatial coordinates output by the visual module | metasurface_adaptive_coding() |
| *"Receiving echo signals and calculating human respiratory rate"* | Cross-environment child process scheduling: Continuously detects the respiratory state of the human body using a respiratory detection algorithm based on received echo signals | respiratory_rate_measure() |

Through the deep coupling of skeletal node perception and physical-layer beamforming, the metaEI-WM, without requiring any wearable medical devices, utilizes RF signal echoes to achieve medical-grade respiratory state monitoring. This highlights the immense potential of intelligent metasurfaces as the core hardware for “integrated sensing and communication (ISAC)”.

## Supplementary Note 2. Fully Automated Acquisition of Metasurface and Signal Source Positions

This supplementary note details the underlying implementation mechanism for the fully automated acquisition of metasurface and signal source positions within the metaEI-WM framework. The primary design objective of this mechanism is to thoroughly “liberate the human brain” by eliminating traditional manual measurements, fiducial marker placement, or cumbersome on-site calibration procedures. The system fundamentally relies on the fully automated Wingman Robot smart car that operates autonomously in indoor scenarios to create a precise spatial semantics (semantically and topologically structured scene) representation through multi-sensor collaboration. For metasurfaces with periodic static structures, the system utilizes an onboard ZED 2i depth camera in conjunction with image matching and geometric pose estimation algorithms for three-dimensional (3D) localization and type recognition. For radio signal sources scattered in practical living environments, the system acquires Channel State Information (CSI) via an Intel 5300 network card mounted on an Ubuntu 14.04 micro-system, and integrates the SpotFi algorithm to perform high-precision passive radio frequency (RF) localization during movement [1]. The fusion of these two data modalities ultimately provides high-fidelity boundary constraints for the digital twin model within a 3D coordinate system, enabling the system to execute autonomous tasks ranging from enhancing non-line-of-sight (NLOS) connections to achieving covert backscatter communication(see Fig. S2).

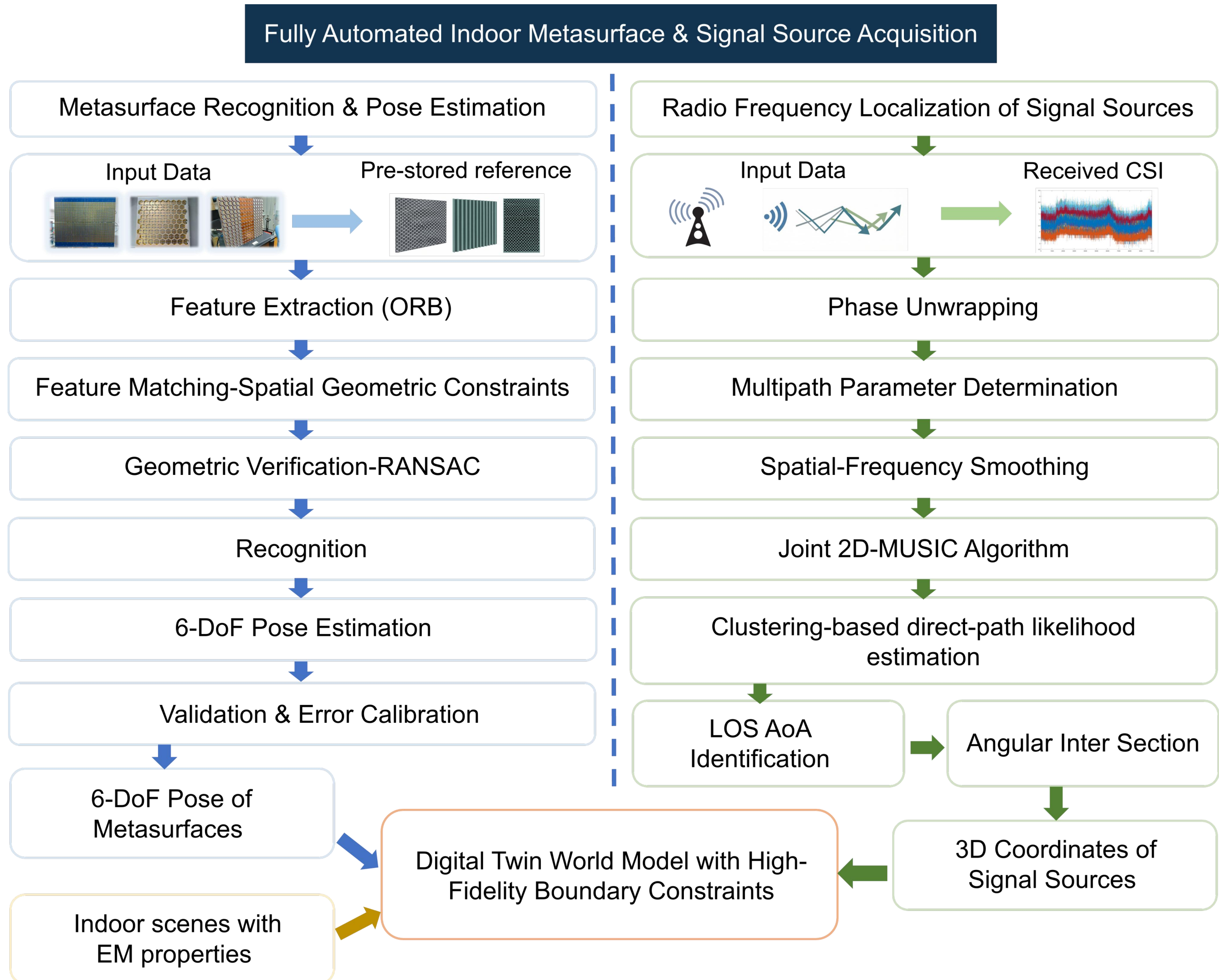

**Fig. S2. Flowchart of fully automated indoor metasurface and signal source spatial acquisition.** This figure illustrates the comprehensive workflow for constructing a high-fidelity digital twin and its corresponding spatial semantics representation of world model. The estimated 6-DoF pose of the metasurfaces, the resolved 3D coordinates of the signal sources, and the indoor scenes with EM properties (detailed in Supplementary Note S3)

are integrated to construct the spatial semantics representation with high-fidelity boundary constraints.

### 1. Metasurface Recognition and Pose Estimation

In this experimental system, a total of three types of metasurfaces with distinct physical characteristics and beam-steering capabilities were employed. Physically, these metasurfaces manifest as vertically positioned rigid planes, whose surfaces are partitioned into a large number of minuscule meta-atom structures with specific periodic arrangements. Given that these three metasurfaces exhibit significant geometric texture features, they can be recognized in real-time through image sequences acquired by the camera mounted on the Wingman Robot or a fixed-view ZED 2i camera. At the optical perception level, the high-resolution binocular RGB lens equipped on the ZED 2i camera provides a wide field of view (FOV) of up to 120 degrees, and driven by an advanced neural depth perception engine, achieves precise spatial depth measurements up to 20 meters. This multimodal data acquisition capability establishes a robust physical data foundation for subsequent 2D image object detection and 3D spatial coordinate mapping. During the initialization phase, the system pre-stores high-resolution reference photographs of the three different metasurfaces from orthogonal perspectives as templates. While the Wingman Robot is in operation, the camera captures grayscale images of the current environment in real-time and applies the ORB algorithm to extract keypoints and local feature descriptors [2]. The ORB algorithm imparts directional information to FAST corners by calculating the intensity centroid of local image patches, and subsequently extracts rotation-invariant BRIEF binary descriptors.

However, the intrinsic property of the metasurfaces—a dense 2D periodic lattice composed of homogenized meta-atoms—introduces a severe challenge to traditional image matching, referred to as "periodic pattern ambiguity". During feature extraction, because the local gradient and texture of every unit cell within the lattice are nearly identical, the descriptor of a meta-atom feature located in the upper left corner of a real-time image may be highly similar in Hamming distance to the descriptor of a meta-atom in the lower right corner of the reference template. Relying solely on a nearest-neighbor matching strategy would result in erroneous matches. Therefore, we introduce spatial geometric constraints: for any given keypoint in the real-time image, the algorithm searches not only for the optimal matching point in the reference template but also for the sub-optimal matching point. If the ratio of the best matching distance to the second-best matching distance exceeds a predefined threshold, it indicates that the feature point lacks global distinctiveness and is highly susceptible to confusion within the periodic structure; consequently, this matching pair is directly discarded.

Subsequently, the remaining candidate matching pairs are subjected to a geometric verification process based on the RANSAC algorithm. Because the metasurface is a perfectly rigid plane, its mapping relationship between the reference image and the real-time perspective image must be strictly constrained by a single 2D projective transformation—namely, the homography matrix. In each iteration, the RANSAC algorithm randomly samples 4 pairs of non-collinear matching points to compute a hypothetical homography matrix, and utilizes this matrix to reproject all other candidate matching points. Points exhibiting projection errors that exceed a pixel threshold are treated as outliers and eliminated. Upon the completion of RANSAC filtering, the system tallies the number of inliers retained by each of the three reference templates. The template with the highest inlier count that also surpasses a confidence threshold is recognized by the system as the metasurface type present in the current field of view. This algorithm, founded on simple photographic geometric matching, improves the robustness of recognition while substantially mitigating computational requirements. Because purely periodic textures may still generate geometrically consistent but physically incorrect matches, the final pose was validated using the known metasurface boundary size and the depth plane measured by the ZED 2i camera.

Upon successfully recognizing the metasurface type and obtaining reliable 2D-2D pixel matching points, the system must transform these into 6-Degree-of-Freedom (6-DoF) pose information within the 3D spatial coordinate system required for the digital twin model. This pose defines the translation vector and rotation matrix of the metasurface relative to the optical center of the Wingman Robot's camera or the fixed-view ZED 2i camera. After calibrating the ZED 2i camera to determine the intrinsic matrix $\mathbf{K}$, and according to the pinhole camera model, the projection relationship mapping a 3D spatial point $\mathbf{P_w} = [X, Y, Z]^T$ to the image pixel plane $\mathbf{p_c} = [u, v]^T$ can be expressed as:

$$s \begin{bmatrix} u \\ v \\ 1 \end{bmatrix} = \mathbf{K}[\mathbf{R}\ \mathbf{t}] \begin{bmatrix} X \\ Y \\ Z \\ 1 \end{bmatrix} \tag{S1}$$

where $s$ denotes the scale factor, $\mathbf{R}$ is a 3×3 rotation matrix, and $\mathbf{t}$ represents a 3×1 translation vector.

Given that the metasurface is defined as a perfectly vertical plane, we define a local metasurface coordinate system whose origin is located at the physical centre of the metasurface, with the metasurface plane set to $Z = 0$. In this configuration, the aforementioned projection equation undergoes dimensionality reduction:

$$\mathrm{s}\begin{bmatrix}\mathrm{u}\\\mathrm{v}\\1\end{bmatrix} = \mathbf{K}[\mathbf{r_1}\ \mathbf{r_2}\ \mathbf{r_3}\ \mathbf{t}]\begin{bmatrix}\mathrm{X}\\\mathrm{Y}\\\mathrm{Z}\\1\end{bmatrix} = \mathbf{K}[\mathbf{r_1}\ \mathbf{r_2}\ \mathbf{t}]\begin{bmatrix}\mathrm{X}\\\mathrm{Y}\\1\end{bmatrix} \tag{S2}$$

Here, $\mathbf{r_1}\ \mathbf{r_2}\ \mathbf{r_3}$ are the orthogonal column vectors of the rotation matrix $\mathbf{R}$. Following this dimensionality reduction, the projection relationship is completely governed by a 3×3 matrix, which corresponds to the homography matrix optimized via RANSAC in the preceding matching stage:

$$\mathbf{H} = \lambda\mathbf{K}[\mathbf{r_1}\ \mathbf{r_2}\ \mathbf{t}] \tag{S3}$$

To extract the actual 3D pose of the metasurface from the homography matrix $H$, the algorithm first left-multiplies by the inverse of the camera's intrinsic matrix, thereby yielding the first two columns of the rotation matrix and the translation vector:

$$\mathbf{M} = \mathbf{K^{-1}H} = [\mathbf{m_1}\ \mathbf{m_2}\ \mathbf{m_3}] = \lambda[\mathbf{r_1}\ \mathbf{r_2}\ \mathbf{t}] \tag{S4}$$

In practical application scenarios, due to image noise and computational inaccuracies, the directly extracted matrix $\mathbf{R_{raw}} = [\mathbf{r_1}\ \mathbf{r_2}\ \mathbf{r_3}]$ may no longer strictly satisfy the orthogonality constraint. Thus, the algorithm employs Singular Value Decomposition (SVD) to approximate the closest valid rotation matrix. Through the aforementioned closed-form algebraic solutions, the Wingman Robot can precisely calculate the spatial coordinates of the metasurface center relative to the camera, alongside its yaw, pitch, and roll angles, utilizing exceptionally rapid photographic matrix operations. The inherent high-precision depth point cloud data from the ZED 2i camera additionally serves as a priori information to conduct depth validation and error calibration on the calculated outcomes. This entire procedure achieves fully automated and seamless operation.

### 2. Radio Frequency Localization of Signal Sources Based on the SpotFi Algorithm

Dynamic radio frequency (RF) signal sources present within the indoor environment constitute another critical class of entities for constructing the world model. In contrast to visible geometric structures such as metasurfaces, the localization of these signal sources necessitates the analysis of invisible electromagnetic wave propagation characteristics throughout the physical space. To this end, metaEI-WM employs the SpotFi algorithm, leveraging Channel State Information (CSI) to extract the Angle of Arrival (AoA) and Time of Flight (ToF) of multipath components, thereby facilitating accurate passive RF localization.

To acquire the foundational CSI data, the Wingman Robot is equipped with a microcomputer running the Ubuntu 14.04 LTS operating system, ensuring perfect compatibility with the Intel Wi-Fi Wireless Link 5300 802.11n MIMO wireless network card with modified firmware. In conjunction with the open-source Linux 802.11n CSI Tool, it records the channel frequency response matrix at the physical layer on a per-packet basis. When configured for a 20 MHz communication bandwidth, the CSI Tool reports the channel matrix corresponding to 30 subcarrier groups out of the 56 Orthogonal Frequency Division Multiplexing (OFDM) subcarriers. Each CSI sample is expressed as a complex number featuring 8-bit precision for both the real and imaginary components, representing the signal attenuation gain and phase shift between a specific pair of transmitting and receiving antennas at a distinct frequency subcarrier.

In indoor environments, signals traveling from a transmission source to the receiving antenna of the Wingman Robot typically do not traverse a single path; rather, they undergo multiple reflections and scattering from walls, ceilings, and metasurfaces. Assuming the signal propagates via $L$ distinct multipaths, the received channel state information $\mathrm{H}_{\mathrm{n,k}}$ for the $k$-th subcarrier and the $n$-th receiving antenna can be modeled as the superposition of signals from all paths:

$$\mathrm{H}_{\mathrm{n,k}} = \sum_{p=1}^{L} \gamma_p e^{-j2\pi f_k \tau_p} e^{-j\pi(n-1)\sin(\theta_p)} \quad \text{(S5)}$$

where $\gamma_p$ denotes the propagation and reflection loss of the $p$-th path; $\tau_p$ represents the time of flight for the $p$-th path; $f_k$ is the center frequency of the $k$-th subcarrier. Because the signal propagates for an identical duration across different frequency subcarriers, it induces distinct frequency-domain phase shifts $e^{-j2\pi f_k \tau_p}$. Furthermore, $\theta_p$ specifies the angle of arrival for the $p$-th path. Given the fixed half-wavelength spatial separation between the three antennas on the Intel 5300 network card, the arrival time of a single plane wave impinging from an angle of $\theta_p$ will exhibit marginal temporal variations across different antennas, consequently generating a spatial-domain phase shift $e^{-j\pi(n-1)\sin(\theta_p)}$. The core mathematical logic of SpotFi revolves around inversely resolving $\tau_p$ and $\theta_p$ for every path by observing a $3 \times 30$ complex matrix constituted by the 3 antennas and 30 subcarriers.

Due to minute frequency deviations between the local clock oscillators at the transmitter and receiver, a sampling frequency offset occurs. Additionally, an unpredictable random time delay arises at the exact moment the receiver's analog-to-digital converter detects an incoming packet and initiates sampling, which is known as the sampling time offset. To restore phase consistency, it is imperative to execute a phase sanitization algorithm utilizing a linear regression mechanism before performing the localization algorithm (for detailed algorithms, please see the referenced paper). Upon obtaining the sanitized and stabilized CSI data, the system advances to the phase of determining multipath parameters.

The central tenet of the SpotFi algorithm is as follows: given that multipath signals present not only AoA-based spatial phase shifts across antennas but also ToF-based frequency phase shifts across subcarriers of different frequencies, the information from the frequency domain can be coupled with spatial domain information to formulate a virtual, large-scale sensor array. The system actualizes this objective through a "spatial-frequency smoothing" technique, conducting sliding window resampling on the 3×30 CSI matrix along both frequency and spatial dimensions. For instance, by grouping the first 15 subcarriers and sliding with a step size of 1 subcarrier per iteration, the raw data can be restructured into a smoothed measurement matrix $\mathbf{X}$ of significantly larger dimensions (e.g., a $30 \times 30$ dimensional matrix). This operation not only restores the full-rank covariance matrix for highly correlated multipath signals but also broadens the detection dimensions. Following the construction of the smoothed matrix $\mathbf{X}$, the system executes the joint 2D-MUSIC algorithm:

(1) Covariance and Subspace Decomposition: Compute the sample covariance matrix $\mathbf{R_{xx}} = \mathbf{XX^H}/\mathbf{N_S}$. Perform an eigenvalue decomposition on $\mathbf{R_{xx}}$, and categorize the eigenvectors corresponding to the smaller eigenvalues into the noise subspace $\mathbf{E_N}$. $\mathbf{N_S}$ is the number of snapshots of the smoothed measurement matrix

(2) 2D Steering Vector Construction: Formulate a theoretical steering vector $\alpha(\theta, \tau)$ that simultaneously encapsulates both angle and time. Its eleme nts comprehensively characterize the theoretical phase that a signal with a specific angle $\theta$ and time $\tau$ should manifest on the $n$-th antenna and the $k$-th subcarrier.

(3) Pseudospectrum Search: Perform a dense grid search within the 2D parameter space of $(\theta, \tau)$ to evaluate the spatial spectrum function:

$$\mathbf{P}(\boldsymbol{\theta}, \boldsymbol{\tau}) = \frac{1}{\boldsymbol{\alpha}(\boldsymbol{\theta}, \boldsymbol{\tau})^{\mathbf{H}} \mathbf{E_N} \mathbf{E_N^H} \boldsymbol{\alpha}(\boldsymbol{\theta}, \boldsymbol{\tau})} \tag{S6}$$

Because the steering vector and the noise subspace must remain orthogonal, the peaks that emerge on the spatial spectrum function represent the genuine angle of arrival and relative time of flight for each existing multipath component.

After isolating all multipath components, the algorithm identifies the component that is most likely to correspond to the direct propagation path between the signal source and the smart car. Because the transmitter and receiver are not tightly time-synchronized, and because packet detection introduces an unknown timing offset, the estimated ToF is relative rather than absolute. The system therefore cannot determine the direct path simply by selecting the component with the shortest ToF. Instead, it follows the direct-path likelihood estimation procedure of the original SpotFi algorithm. The AoA–ToF pairs estimated from multiple packets are clustered in the joint $(\theta_p, \tau_p)$ domain, and each cluster is assigned a likelihood score according to its temporal consistency, statistical compactness and support across packets. This procedure exploits the fact that the direct path generally exhibits more consistent AoA–ToF estimates than many reflected paths, while avoiding the assumption that the strongest or shortest-delay component must be the direct path. The cluster with the highest direct-path likelihood is selected as the most probable direct-path component, and its AoA is used as the bearing measurement of the target signal source. As the car navigates the space, AoA rays obtained from multiple known robot poses are then intersected to estimate the horizontal position of the signal source. When a full 3D coordinate is required, the height coordinate is supplied by prior installation information, an additional elevation-angle estimate or an independent geometric constraint.

**Supplementary Note 3. Detailed Implementation of Heterogeneous Perception & Representation Module**

This supplementary note elaborates on a highly integrated pipeline for the autonomous reconstruction of physical-electromagnetic twins (see Fig. 2). The metaEI-WM, using an intelligent vehicle as its hardware carrier (see Fig. S3), achieves a thorough parsing of unfamiliar indoor scenes through the seamless integration of multiple algorithmic modules. The entire processing pipeline follows a strict logical progression. First, the metaEI-WM invokes the TARE (Technologies for Autonomous Robot Exploration) hierarchical framework [3] to address autonomous navigation and blind-spot-free path planning in unknown spaces. While the vehicle is moving, the R3LIVE tightly-coupled state estimation system [4] is utilized to acquire a real-time 3D dense point cloud with high-precision RGB colors. Subsequently, this point cloud is fed into the SpatialLM [5] module, which is based on a large language model architecture. This module discards traditional heuristic clustering and directly extracts 3D bounding boxes and semantic entities in the form of structured code (Python Dataclass). Then, incorporating the "polygon set prediction" concept of RoomFormer [6], the system transforms the discrete semantic entities into spaces of different functional zones with clear boundaries (e.g., bedrooms, kitchens). Finally, the metaEI-WM maps the aforementioned semantic and geometric information onto the preset ITU-R P.2040 standard electromagnetic material library, thereby assigning physical properties such as permittivity to the geometric entities.

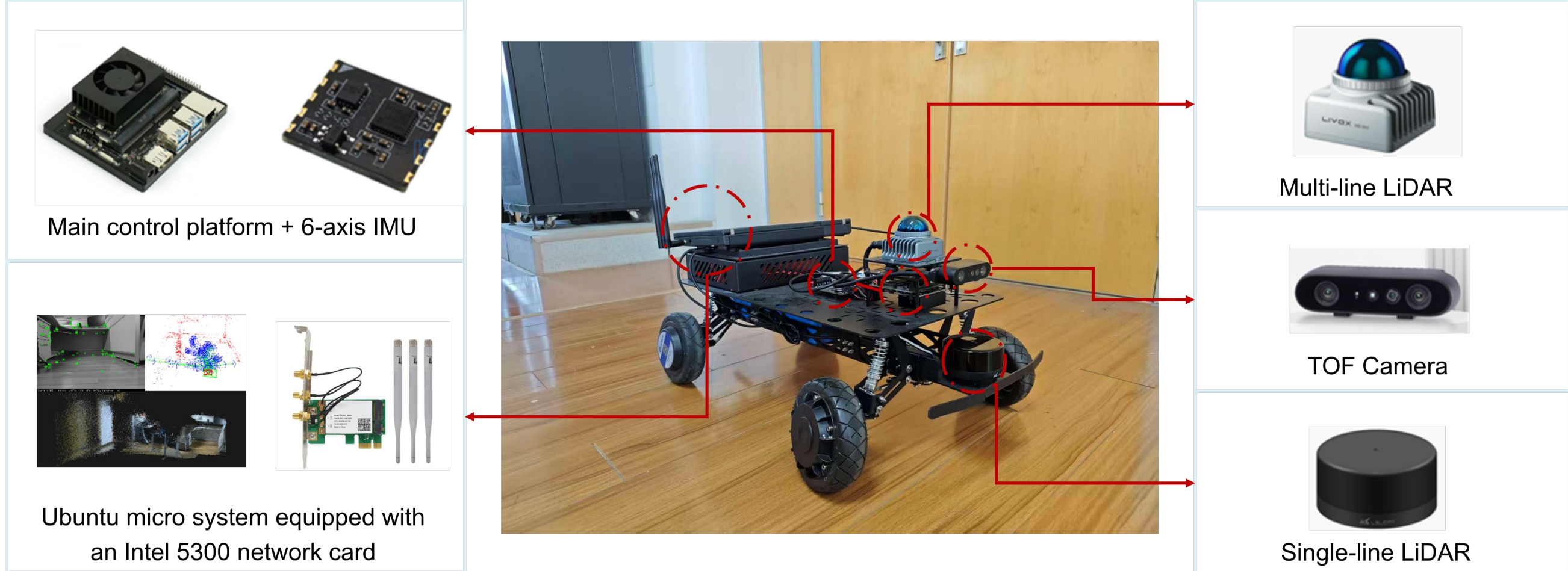

**Fig. S3. An integrated intelligent robot with heterogeneous perception and computing platform.** The computing and sensor stack of this intelligent robot consists of the NVIDIA Jetson Orin NX main control platform, the STM32 MCU expansion board (integrating 6-axis IMU), and a comprehensive multi-sensor array. The array includes multi-line lidars, TOF cameras, and single-line lidars. Additionally, to capture and perceive the underlying physical signals, the platform is specially equipped with an Ubuntu microsystem equipped with an Intel 5300

network card, which is specifically used for real-time acquisition of WiFi CSI signals and supports indoor source localization.

This series of cross-modal transformations, from low-level sensor data to high-level physical semantics, completely breaks down the barrier between autonomous robot mapping and wireless communication channel simulation. This provides a complete, high-fidelity prior physical engine for acquiring 2D and 3D spatial semantics, which serve as the foundation for subsequent ray-tracing simulations to construct the world model.

## 1. Autonomous Space Exploration and Data Collection Based on the TARE Hierarchical Framework

The fundamental prerequisite for generating a digital twin is the ability of an autonomous mobile robot to conduct highly efficient, exhaustive spatial traversal and sensor data collection in entirely unstructured indoor environments. To achieve this, the data acquisition module of metaEI-WM deploys the TARE hierarchical autonomous exploration framework. The essence of this approach is inspired by human cognitive mechanisms in unknown environments, logically decoupling a large-scale, complex space into "local" and "global" tiers. Macroscopically, the system maintains a sparse topological graph that captures merely the connectivity among coarse subspaces; this serves to guide the robot toward remote, unmapped frontiers in the later stages of exploration. Microscopically, within the robot's active local planning horizon, the system maintains a high-resolution, dense environment representation that includes geometry and material information.

Within the local planning horizon $\boldsymbol{\mathcal{H}}$, the core of the exploration problem is modeled as finding the shortest kinematically feasible path, ensuring that this path covers all surfaces currently discovered by exploration frontiers but not yet fully observed. Let the robot sensor's current viewpoint be $\mathbf{v}_{\text{current}}$, the system uniformly samples a set of candidate viewpoints $\mathbf{V}$ within the local configuration space $\boldsymbol{C}_{\boldsymbol{trav}}^{\mathcal{H}}$. For any unobserved surface patch in the environment with a center point coordinate $\mathbf{p_s}$ and a normal vector $\mathbf{n_s}$, the system must determine whether it can be covered by a candidate viewpoint $\mathbf{v}$ located at $\mathbf{p_v}$. The patch is considered successfully covered if and only if the following two core geometric constraints are met:

$$|\mathbf{p_s} - \mathbf{p_v}| \leq D \tag{S7}$$

$$\frac{\mathbf{n_s} \cdot (\mathbf{p_s} - \mathbf{p_v})}{|\mathbf{n_s}||\mathbf{p_s} - \mathbf{p_v}|} \geq \mathbf{T} \tag{S8}$$

In the core logical formulas above, $D$ represents the maximum effective detection distance of the sensor, restricting the viewpoint from being too far from the target; while $\mathbf{T}$ is the threshold parameter for the predicted viewing angle.

Due to the overlapping fields of view from different viewpoints, the viewpoint reward function (i.e., the unknown surface area $A_v$ that can be covered) mathematically exhibits typical submodularity. That is, for any two viewpoint subsets $\mathbf{V'} \subseteq \mathbf{V}$ and a single new viewpoint $\mathbf{v}$, its marginal coverage benefit satisfies the diminishing return property:

$$a(\mathbf{V}' \cup \mathbf{v}) - a(\mathbf{V}') \geq a(\mathbf{V} \cup \mathbf{v}) - a(\mathbf{V}) \tag{S9}$$

Leveraging this diminishing return property, the system avoids time-consuming global combinatorial optimization and instead adopts a greedy iterative sampling strategy with a priority queue. Each time, it extracts only the viewpoint that brings the maximum marginal benefit and rapidly applies a decay update to the reward values of the remaining viewpoints in the queue. This allows the system to converge on a near-optimal set of covering viewpoints within an extremely short computational cycle.

After determining the local viewpoints, a smooth path passing through these viewpoints needs to be generated. TARE transforms this problem into an Asymmetric Traveling Salesperson Problem (ATSP) with penalty terms. Assuming the generated piecewise path has a total of $\widetilde{n}$ segments, the length of the $j$-th segment is $l_j$, and the penalty factor for stopping and turning at breakpoints is $p$, the cost function $c^{'}_{smooth}$ for the local smooth path is defined as:

$$c^{'}_{smooth} = \sum_{j=1}^{\widetilde{n}} l_j + p(\widetilde{n} - 1) \tag{S10}$$

After solving for the traversal sequence with the lowest cost using heuristic algorithms (such as the Lin-Kernighan algorithm), marginalized trajectory optimization methods are further utilized to fit cubic splines between nodes. This process ensures collision clearance while guaranteeing the continuity of the vehicle's velocity and acceleration. Ultimately, under the guidance of TARE, the intelligent vehicle navigates through unknown indoor scenes with an extremely smooth and efficient posture, laying a perfect kinematic foundation for subsequent real-time data collection.

**2. High-Fidelity Geometric Reconstruction based on the $R^3$LIVE Framework**

As the intelligent vehicle achieves comprehensive coverage of the indoor space driven by the TARE framework, its onboard multi-modal sensors (LiDAR, IMU, visual cameras) capture environmental data at a high frequency. For electromagnetic ray-tracing simulations, a pure geometric point cloud is far from sufficient. Electromagnetic waves are highly sensitive to the material of object surfaces, and the inference of materials relies heavily on high-fidelity visual texture features. Therefore, the system adopts the $R^3$LIVE tightly-coupled state estimation and mapping framework. It consists of two synergistically operating subsystems: a LiDAR-Inertial Odometry (LIO) and a Visual-Inertial Odometry (VIO), which jointly optimize a state vector within a unified Error-State Iterated Kalman Filter (ESIKF) framework. This state vector $\boldsymbol{x} \in \mathbb{R}^{\mathbf{29}}$ includes the IMU's pose, velocity, bias, global gravity acceleration, as well as the extrinsic parameters between the camera and IMU and the camera's intrinsic parameters. The dense point cloud with precise RGB information output by this framework serves as the direct data source for subsequent semantic segmentation by large models and EM material allocation.

As the autonomous mobile robot achieves comprehensive coverage of the indoor space guided by the TARE framework, its onboard multimodal sensors (LiDAR, IMU, and visual cameras) synchronously acquire environmental data at a high frequency. For electromagnetic wave ray-tracing simulations, a purely geometric point cloud is inherently insufficient; electromagnetic waves are highly sensitive to object surface materials, the inference of which relies heavily on high-fidelity visual texture features. Consequently, parallel to the execution of TARE, metaEI-WM deploys the R3LIVE tightly-coupled state estimation and mapping framework. This framework comprises two synergistically operating subsystems—LiDAR-inertial odometry (LIO) and visual-inertial odometry (VIO)—which jointly optimize a state vector within a unified Error-State Iterated Kalman Filter (ESIKF) architecture. This state vector $\boldsymbol{x} \in \mathbb{R}^{\mathbf{29}}$ encapsulates the IMU pose, velocity, and biases, the global gravity acceleration, as well as the extrinsic parameters between the camera and the IMU, alongside the camera intrinsics. The dense point cloud enriched with precise RGB information, generated by this framework, serves as the direct data source for subsequent large model-based semantic segmentation and electromagnetic material assignment.

The LIO subsystem is responsible for building the "skeleton" of the map. Each point in the map contains not only strict 3D coordinates but also maintains the covariance matrix of its coordinates, representing the uncertainty of its spatial position. For continuous input LiDAR scans, the LIO uses the IMU's backward propagation to accurately compensate for motion distortion within the scan frame. Subsequently, the ESIKF minimizes the point-to-plane residuals between the newly input point cloud and the global voxel map. Once the pose estimation converges, the new frame of the point cloud is approved and registered into the global voxel map, forming an extremely dense and accurate geometric structural skeleton.

The VIO subsystem is responsible for constructing the "skin" of the map, rendering the color and texture features of the point cloud. It directly fuses visual data by minimizing the frame-to-map photometric error. Its core logic is that for a 3D point already existing in the global map, regardless of how the camera moves, the color of its projected point on the current camera image should be highly consistent with the prior color saved for that 3D point in the map. Let the $s$-th point recorded in the global map be $P_s$, with its prior color being $c_s$. Given the current camera pose, when this point is projected onto the current image frame $I_k$, the observed color value $\gamma_s$ can be measured via non-linear interpolation. Thus, the frame-to-map photometric error residual function is defined as:

$$o(\widetilde{x_k}, p_s^G, c_s) = c_s - \gamma_s \tag{S11}$$

This formula is the essence of the VIO state estimation. By substituting this residual model into the ESIKF framework, the system does not merely "color" the map; instead, it utilizes pixel-level photometric differences to reversely correct and optimize the system's own state estimation. This mechanism enables the metaEI-WM system to rely on visual textures to maintain localization when encountering geometric degradation, and to rely on LiDAR for stability when facing visual degradation, demonstrating extremely high robustness.

Meanwhile, to overcome the random walk noise $\eta_{c_s}$ caused by environmental lighting changes, after the pose converges, the system utilizes Bayesian update rules to refresh the RGB color and its variance for each point in the map in real-time:

$$\Sigma_{n\widetilde{c_s}} = \left(\left(\Sigma_{nc_s} + \sigma_s^2 \cdot \Delta t_{c_s}\right)^{-1} + \Sigma_{n\gamma_s}^{-1}\right)^{-1} \tag{S12}$$

$$\widetilde{c_s} = \Sigma_{n\widetilde{c_s}}\left(\left(\Sigma_{nc_s} + \sigma_s^2 \cdot \Delta t_{c_s}\right)^{-1} c_s + \Sigma_{n\gamma_s}^{-1}\gamma_s\right) \tag{S13}$$

Here, $\Sigma_{nc_s}$ is the prior state, the updated $\Sigma_{n\widetilde{c_s}}$ represents the posterior state after fusing the current observation, $\Sigma_{n\gamma_s}$ is the noise covariance of the current color observation, and $\sigma_s^2 \cdot \Delta t_{c_s}$ represents the variance inflation injected as time elapses. When a point has not been observed for a long time, the system gradually reduces its trust in its old color; whereas when the point is continuously observed across multiple frames in a short period, the high-confidence data will dominate the color fusion. Through this Bayesian mechanism, $R^3$LIVE not only outputs a geometric skeleton with sub-centimeter accuracy but also endows the map with extremely smooth visual details that resist lighting interference. This step is crucial because subsequent deep learning models heavily rely on these high-quality RGB features for the semantic segmentation and extraction of materials.

**3. Structured Extraction of Semantic Entities via SpatialLM**

Although the $R^3$LIVE system generates a dense, colored point cloud map, from the perspective of computational and communication simulation engines, the raw 3D point cloud data—comprising solely spatial coordinates (XYZ) and RGB color values—inherently lacks semantic information. Consequently, it cannot be directly applied to material analysis and propagation loss calculations within wireless channels. Conventional methods that process point clouds via voxelization or heuristic clustering are highly susceptible to the degradation of spatial resolution and the loss of fine-grained features. To overcome these limitations, metaEI-WM leverages the SpatialLM framework [5], which is founded upon a Large Language Model (LLM) architecture, to transform the unstructured point clouds yielded by the $R^3$LIVE module into a highly structured semantic description protocol.

SpatialLM strictly follows the standard "Encoder-MLP-LLM" multi-modal architecture. At the encoding front-end, the system employs the decoder-free Point Transformer V3 as the 3D point cloud backbone network. Assuming an input point cloud $P \in \mathbb{R}^{N\times 6}$ (containing 3D coordinates and RGB colors) from $R^3$LIVE, Point Transformer V3 directly extracts local and global joint geometry-color features on the raw point set through its unique multi-scale deformable attention mechanism. After encoding, the massive point cloud is highly compressed into a set of compact visual tokens $\mathbf{F} \in \mathbb{R}^{K\times D}$, where $K \ll N$ is the number of visual features and $D$ is the feature dimension. To enable the LLM to understand these 3D features, the system uses a projection matrix consisting of a two-layer multi-layer perceptron to forcefully align and map the visual tokens $F$ from the 3D geometric space into the text embedding space used for the LLM's pre-training.

Upon receiving the mapped multi-modal features and user-provided text prompts (e.g., "Detect walls, doors, windows, boxes"), the LLM directly generates pure text code describing the scene through its powerful autoregressive mechanism. This innovative "scene-as-code" paradigm defines a set of universal Python Dataclass structures.

In this way, the originally complex tasks of 3D bounding box regression and instance segmentation are elegantly transformed into a "next-word prediction" task in natural language processing. The LLM directly predicts oriented 3D bounding boxes containing architectural elements like "walls, doors, and windows," as well as physical entities like furniture and appliances. This mechanism endows the model with strong generalization capabilities and zero-shot reasoning potential, maintaining physical consistency across different point cloud input sources (such as depth cameras, reconstructed meshes, or LiDAR). The Table Supplementary Note 4 compares the differences between traditional methods and the SpatialLM architecture, highlighting the engineering advantages of applying SpatialLM in the metaEI-WM.

Through the deep parsing by SpatialLM, the raw point cloud is stripped away, replaced by a series of oriented object bounding boxes and architectural planes with clear azimuths, length-width-height dimensions, and precise semantic labels (e.g., "wooden desk, metal cabinet, concrete wall"). At this point, the physical world has completed its preliminary physical abstraction.

**Table S4** Comparison of engineering advantages between traditional 3D detection networks and the SpatialLM architecture

| **Feature Dimension** | **Traditional 3D Detec-tion Networks** | **SpatialLM Large Lang-uage Model Architec-ture** | **Engineering Advantages in metaEI-WM** |
|---|---|---|---|
| **Output Format** | Discrete tensors and confidence matrices that are difficult for large models to interpret | Structured executable text code | Extremely high interpretability; generated parameters can be directly parsed by automated scripts and injected into the RF simulation engine |
| **Category Expansion** | Adding new objects requires modifying the network classification head and collecting massive data for retraining | Possesses powerful zero-shot or few-shot generalization capabilities based on prompt words | New electromagnetic reflector categories can be added at any time via natural language instructions |

| **Spatial Reasoning** | Relies solely on local geometric features, lacking global common-sense connections between entities | Inherits the profound common-sense reasoning capabilities of LLMs with tens of billions of parameters | Automatically infers physical occlusion and support relationships of objects, providing underlying physical logic support for ray tracing |
|---|---|---|---|

**4. Macroscopic Topological Space Segmentation and Automated Mapping of the EM Material Objects**

Although metaEI-WM successfully identifies discrete structural components within the environment by leveraging the SpatialLM module, an indoor space is fundamentally more than a mere aggregation of physical entities. For metaEI-WM to fully interpret natural language instructions from users, it is imperative to reconstruct a seamless indoor floorplan and explicitly delineate the functional zones corresponding to individual "rooms." To achieve this, metaEI-WM employs the RoomFormer module, effectively reformulating the floorplan reconstruction process as a single-stage, "direct polygon set prediction" task.

Since each room contains a different number of corner points, the system first projects the 3D features acquired by $R^3$LIVE along the direction of gravity into a 2D bird's-eye view density map, which is then fed into a Transformer decoder. The decoder utilizes a two-level query mechanism for iterative reasoning: first, room-level queries are responsible for determining the existence probabilities of a maximum of $M$ potential rooms; second, corner-level queries iteratively regress the coordinates of up to $N$ ordered vertices $\mathbf{p}_\mathbf{n}^\mathbf{m}$ and their validity flags $c_n^m \epsilon \{0,1\}$ within each room through the self-attention and cross-attention mechanisms of the Transformer decoder. Because the system outputs ordered vertex coordinates with strong sequential features, sequentially connecting these valid vertices naturally forms a perfectly closed room polygon contour. The set of polygons $\hat{\mathbf{S}} = \{\hat{\mathbf{V}}_\mathbf{m}\}_{\mathrm{m}=1}^{\mathrm{M}}$ predicted by the model and the true physical structure set are unordered, where each polygon represents a room $\hat{\mathbf{V}}_\mathbf{m} = (\hat{\mathbf{v}}_\mathbf{1}^\mathbf{m}, \hat{\mathbf{v}}_\mathbf{2}^\mathbf{m}, \ldots, \hat{\mathbf{v}}_\mathbf{N}^\mathbf{m},)$.

To achieve alignment between the predicted polygons and the true physical structures, the system employs bipartite graph matching based on the Hungarian algorithm during reconstruction to find the permutation $\hat{\sigma}$ that minimizes the global matching cost $D$:

$$\widehat{\sigma} = \arg\min \sum_{m=1}^{M} D(\mathbf{V_m}, \widehat{\mathbf{V}}_{\boldsymbol{\sigma}(\mathbf{m})}) \tag{S14}$$

Here, the calculation of the matching cost $D$ comprehensively considers the classification cross-entropy loss $\mathcal{L}_{\mathbf{cls}}$ of the vertex validity labels, the coordinate regression loss $\mathcal{L}_{\mathbf{coord}}$, and the polygon Dice loss $\mathcal{L}_{\mathbf{ras}}$ based on differentiable rasterization.

The generated ordered vertex sequence naturally closes to form the polygon contour of the room. Furthermore, the system spatially associates the oriented object bounding boxes extracted by metaEI-WM with the closed polygons. Leveraging the LLM's contextual abilities, heuristic corrections are performed: if a closed region contains “bed, wardrobe, and nightstand,” the polygon topological node is labeled as “Bedroom”; if it contains a “toilet and sink,” it is labeled as “Bathroom”. This node graph with macroscopic semantics greatly enhances the system's decision-making capabilities, enabling the metaEI-WM to understand high-level natural language instructions involving room topological semantics, substantially improving the autonomy of environmental perception and decision-making.

Through the aforementioned steps, the metaEI-WM has successfully reconstructed the initially chaotic and unordered physical space into a comprehensive spatial semantics representation possessing fine geometric coordinates, closed topological regions, and rich entity semantics. However, for this model to truly play a role in intelligent radio environments and execute channel simulations and IMS adjustments under extremely low latency, the pure geometric model must be endowed with real electromagnetic physical properties. Benefiting from the high-fidelity structured semantic information generated by the system's front-end, the system achieves end-to-end fully automated allocation of electromagnetic material parameters.

The system embeds an “electromagnetic material library” strictly referencing the International Telecommunication Union (ITU-R) P.2040 standard. The ITU-R P.2040 standard is a recommended standard for the radio frequency propagation characteristics of building materials in specific frequency bands, providing a series of empirical formulas to calculate the complex permittivity $\varepsilon_r$ and conductivity $\sigma$ of different materials at different frequencies $f$. An automated allocation script iterates through every semantic cluster within the structured text output by SpatialLM. Through preset dictionary mapping relationships, it automatically substitutes the target operating frequency for calculation and executes physical property injection, thereby replacing the original structured bounding boxes with triangular mesh entities possessing accurate electromagnetic properties. The Table S5 shows examples of parameter calculations for some frequently occurring indoor semantic labels mapped to the 5 GHz band after extraction.

**Table S5** Electromagnetic material parameter calculation examples for indoor semantic entities mapped to the 5 GHz ban

| Semantic Entity Label | Standard Material Category | Relative Permittivity $\varepsilon_r$ (Real Part) | Conductivity $\sigma$ (S/m) | Ray-Tracing Polygon Modeling |
|---|---|---|---|---|
| **Window Glass** | Glass | 4.5 | 0.008 | |
| **Wall / Ceiling** | Concrete | 5.31 | 0.12 | |
| **Metal Cabinet** | Metal | 1 | 1e7 | |
| **Wooden Desk** | Wood | 1.99 | 0.026 | |
| **Wooden Door** | Wood | 1.99 | 0.026 | |
| **Chair** | Plastic | 2.5 | 0.005 | |
| **Sofa** | Fabric / Leather | 1.5 | 0.005 | |

<table>
<tr><td>Bed</td><td>Fabric / Foam</td><td>1.2</td><td>0.002</td><td></td></tr>
<tr><td>Toilet</td><td>Ceramic</td><td>5.5</td><td>0.015</td><td></td></tr>
</table>

After completing the mapping, every 3D semantic bounding box in the system is replaced with an intersectable entity possessing electromagnetic properties. These entities not only have sub-centimeter absolute coordinates derived from $R^3$LIVE but also carry precise permittivity and conductivity that comply with international standards.

Subsequently, the system automatically packages and integrates this multi-dimensional, multi-modal fused dataset, outputting it as a standardized OBJ file with independent material label groups, which interfaces with ray-tracing simulators that support GPU parallel acceleration. This coherent and smooth automated construction pipeline truly builds a solid bridge among the physical world, machine vision, large language model understanding, and digital electromagnetic channel simulation. It provides a powerful digital foundation—requiring zero manual intervention, possessing macroscopic topological awareness, and complete with physical parameters—for future intelligent wireless environment optimization and the simulation of dynamic electromagnetic wave reflection and diffraction processes.

## Supplementary Note 4. Detailed Explanation of World-Model EM Dynamics Prediction of IMS-Assisted Indoor Channels

### 1. Indoor Electromagnetic (EM) Scenario and Ray Tracing Engine Architecture

To precisely simulate the reconstructed indoor electromagnetic dynamics within the digital domain, metaEI-WM integrates a highly efficient, GPU-accelerated EM ray-tracing engine based on the Ray Launching method. This engine directly employs the spatial semantics representation—generated by the preceding modules and annotated with semantic and electromagnetic material labels—as its simulation environment. Drawing upon the paradigm of multi-level voxelization [7], the engine achieves deep optimization by leveraging the NVIDIA CUDA OptiX ray-tracing framework. This acceleration architecture deconstructs the indoor physical space into three logical tiers:

- Scene-level macro-voxels: At the highest level, the entire indoor space is discretized into a low-resolution uniform voxel grid. This level consumes minimal VRAM and primarily provides macroscopic spatial accessibility information; each macro-voxel records the shortest stepping distance to the nearest voxel containing geometric entities.
- High-resolution sub-voxels: For macro-voxels containing environmental geometric details, a divisor factor $D_u$ is used to further partition the space, forming a high-resolution sub-voxel grid. This approximates the boundaries of irregular objects and reduces subsequent invalid computations.
- Mesh Intersectable Entities (MIEs): At the lowest level, high-resolution sub-voxels containing geometric entities are ultimately partitioned using a divisor factor $D_v$, keeping the error within an acceptable range. In this part, the OptiX framework leverages its built-in RT Cores hardware acceleration units to construct low-level Axis-Aligned Bounding Boxes (AABB) for each MIE, and encapsulates the vertices and material indices of the triangular meshes into the ray tracing pipeline.

During the Ray Traversal process, the ray first executes a 3D Digital Differential Analyzer (3D-DDA) algorithm based on 3D texture caching within the low-resolution voxel grid for rapid stepping. Assuming the parametric equation of the ray is $\mathbf{V_{new}} = \mathbf{V_{pos}} + \mathbf{R_{dir}} \cdot \mathbf{S_{total}}$, where $\mathbf{V_{pos}}$ is the current coordinate, $\mathbf{R_{dir}}$ is the ray's direction vector, and $\mathbf{S_{total}}$ is the stepping scalar. The algorithm reads the stepping distance factor $A_{dist}$ (representing the safe clearance distance to the nearest voxel containing an MIE entity) indexed by the current voxel position from memory, and determines the final stepping length through highly parallel vector operations:

(1) Calculate the ray parameter increment vector required to advance one unit coordinate length along each axis, where $\epsilon_z$ is a small bias to prevent division by zero:

$$\mathbf{S_{unit}} = \frac{1}{\max\left(abs(\mathbf{R_{dir}}), \epsilon_z\right)} \tag{S15}$$

(2) Calculate the physical distance from the current coordinate to the next voxel boundary, and obtain the step size $\mathbf{S_{next}}$ to the next boundary via the Hadamard product:

$$\mathbf{S_{next}} = D_{next} \odot \mathbf{S_{unit}} \tag{S16}$$

(3) Combining the voxel clearance distance $A_{dist}$, calculate the total step vector $\mathbf{T}$ required to traverse this distance:

$$\mathbf{T} = \mathbf{S_{next}} + \mathbf{S_{unit}} \cdot (A_{dist} - 1) \tag{S17}$$

(4) Utilize bitwise logic to determine the minimum value index $X$ of the x, y, z components in $T$ (i.e., the boundary the ray penetrates first), finally yielding the precise total stepping length $S_{total}$:

$$\mathbf{S_{total}} = T \cdot \mathbf{X} + \boldsymbol{\epsilon} \tag{S18}$$

This stepping mechanism, based on integer logic and fast vector multiply-add operations, is highly suited to the Single Instruction, Multiple Threads (SIMT) architecture of GPUs, eliminates complex tree-like branch predictions, and significantly reduces ray penetration time in large-scale spaces. When the ray steps into a sub-voxel confirmed to contain an MIE, it triggers OptiX's low-level, high-precision Möller-Trumbore triangle intersection kernel. This algorithm directly determines the intersection between the ray $\boldsymbol{o} + t\boldsymbol{d}$ and the triangular face containing vertices $\mathbf{v_0}, \mathbf{v_1}, \mathbf{v_2}$:

$$\begin{bmatrix} t \\ u \\ v \end{bmatrix} = \frac{1}{\det(M)} \begin{bmatrix} (\boldsymbol{s} \times \boldsymbol{e_1}) \cdot \boldsymbol{e_2} \\ (\boldsymbol{d} \times \boldsymbol{e_2}) \cdot \boldsymbol{s} \\ (\boldsymbol{s} \times \boldsymbol{e_1}) \cdot \boldsymbol{d} \end{bmatrix} \tag{S19}$$

where the edge vectors are $\boldsymbol{e_1} = \boldsymbol{v_1} - \boldsymbol{v_0}$ and $\boldsymbol{e_2} = \boldsymbol{v_2} - \boldsymbol{v_0}$, the vector from the ray to the vertex is $\boldsymbol{s} = \boldsymbol{o} - \boldsymbol{v_0}$, and the determinant is $\det(\boldsymbol{M}) = (\boldsymbol{d} \times \boldsymbol{e_2}) \cdot \boldsymbol{e_1}$. NVIDIA RT Cores perform extremely deep hardware pipeline optimization for the calculation of this determinant and cross products, massively improving the clock cycle efficiency of the intersection tests.

By deploying the ray-tracing engine, metaEI-WM executes pure geometric ray tracking throughout the indoor physical space, ultimately yielding a comprehensive set of valid ray paths. For every ray that successfully arrives at the receiver, metaEI-WM records not merely its multipath topological sequence from the source to the observation point, but also precisely outputs the 3D coordinates of all intersection points, the angles of incidence, and the semantic and material index labels bound to the intercepted geometric entities. This a priori data, derived strictly from the geometric dimension, serves directly as the computational input for the subsequent electromagnetic physics surrogate model.

## 2. Electromagnetic Physics Proxy Model and Loss Evaluation

Based on the deterministic geometric paths and material indices output by the preceding ray tracing engine, this module aims to map the abstract ray topology to specific electromagnetic physical losses. metaEI-WM evaluates the electromagnetic attenuation of each ray based on the far-field approximation of Maxwell's equations and high-frequency asymptotic solving methods. The semantic labels of the environmental mesh are mapped to the frequency-dependent parameters of the ITU-R P.2040 specification. The complex permittivity $\eta(f)$ encompassing the real part describing polarization and the conductivity part describing absorption loss, is defined as:

$$\eta(f) = \left(a \cdot f^{b}\right) - j\left(\frac{c \cdot f^{d}}{2\pi f \epsilon_0}\right) \tag{S20}$$

where $f$ is the carrier frequency, and the coefficients $a, b, c, d$ ensure the high fidelity of this dispersion model in the 5G and even millimeter-wave bands.

Regarding to the reflection of electromagnetic waves at different medium interfaces, the system separately calculates the Fresnel reflection coefficients $\Gamma$ for Transverse Electric (TE) and Transverse Magnetic (TM) waves. Let the incident angle be $\theta_i$:

$$\Gamma_{TE}(\theta_i, \eta) = \frac{cos\ \theta_i - \sqrt{\eta - \sin^2\theta_i}}{\cos\theta_i + \sqrt{\eta - \sin^2\theta_i}} \tag{S21}$$

$$\Gamma_{TM}(\theta_i, \eta) = \frac{\eta cos\ \theta_i - \sqrt{\eta - \sin^2\theta_i}}{\eta\cos\theta_i + \sqrt{\eta - \sin^2\theta_i}} \tag{S22}$$

When electromagnetic waves are in Line-of-Sight (LoS) obstructed areas, diffraction occurring from the sharp edges of furniture or doors and windows can contribute non-negligible signal coverage energy. The engine employs the Keller cone model; when an incident ray hits a mesh boundary with edge semantic topology, it excites a family of diffracted rays located on a conical surface. According to the Uniform Theory of Diffraction (UTD), the amplitude of the diffracted electric field $E_d$ is determined by the product of the incident field $E_i$ and the diffraction coefficient tensor $D$, supplemented by a distance spreading factor:

$$E_d = E_i \cdot D(\alpha, \theta_{inc}) \cdot \sqrt{\frac{\rho}{s(\rho + s)}} e^{jk_0 s} \tag{S23}$$

where $\rho$ represents the principal radius of curvature of the wavefront divergence, and $s$ is the distance from the diffraction point to the receiver. By rigorously calculating the Fresnel transition function, the system guarantees the continuity and energy conservation of the diffracted field at the boundary of light and dark regions, thereby ensuring that the world model can capture weak multipath echoes even in severely occluded blind zones.

**3. IMS Cascade Response and World-model Channel Prediction**

Following the aforementioned physical computations, the system successfully extracts the precise amplitude attenuation and phase shifts for all static Line-of-Sight (LoS) and Non-Line-of-Sight (NLoS) multipath components within the environment. Building upon this foundation, metaEI-WM incorporates IMS into the propagation link to construct a full-space channel model parameterized by dynamic control variables. The system tracks three core path topologies: unobstructed direct paths, environmental multipaths, and RIS-cascaded multipaths. When a ray intersects an RIS entity, it circumvents conventional laws of reflection and instead triggers a generalized reflection kernel. After acquiring the initial quantized phase distribution arriving at the RIS aperture from the transmitter, the system dynamically computes the anomalous reflection angle and the abrupt phase response of localized regions according to the active phase coding matrix. Through the coherent superposition of complex electric fields within the CUDA kernel, the system rapidly computes the Channel Impulse Response (CIR), furnishing a physical a priori model for subsequent coding optimization.

Aligned with the 5G standardized channel model 3GPP TR 38.901, metaEI-WM adopts an RIS-assisted deterministic channel modeling methodology. This approach comprehensively assimilates the multipath parameters extracted via ray tracing, formulating the channel response as a joint transfer matrix that encapsulates both geometric parameters and the RIS coding state. This paradigm, which couples geometrical optics with parallel computing, facilitates the real-time output of physically interpretable and predictable signal strengths. Consequently, it effectively supersedes the offline parameter table lookups and statistical cluster generation inherent in conventional empirical models, thereby exhaustively capturing the deterministic electromagnetic signatures of the indoor scenario.

In a basic communication system with a single Tx and Rx based on IMS, let $\psi_l = \alpha_l e^{j\omega_l}$ represent the programmable reflection coefficient of the IMS element $U_l$, where $l \in [1, \cdots, L = M \times N]$, $P_T$ is the Tx transmit power, $G_T$ and $G_R$ are the gains towards the wave direction, $G$ is the gain of the IMS element, $\lambda$ is the wavelength, $a_l$ and $b_l$ are the distances from element $U_l$ to the transmitting and receiving antennas respectively, and $\zeta$ is the loss of each path interacting with the environment, calculated from the previous section. According to literature [8], the received power via the $l$-th element to the receiving antenna is:

$$P_R^{\mathrm{l}} = P_T \frac{\zeta G_T G_R G_t G_r F_l^{combine} \lambda^4}{(4\pi)^4 a_l^2 b_l^2} \tag{S24}$$

And $F_l^{combine} = F^T F^t F^r F^R$. Unifying the effects of radiation pattern, gain, and other losses using $G_e^2$ as a substitute, and due to the scattering characteristics of the IMS elements, the total path attenuation can be split into the product of two independent attenuations:

$$Q_l = \frac{P_R^{\mathrm{l}}}{P_T} = Q_l^1 \times Q_l^2 = \frac{G_e \lambda^2}{(4\pi)^2 a_l^2} \times \frac{G_e \lambda^2}{(4\pi)^2 b_l^2} \tag{S25}$$

Accordingly, the received discrete-time baseband signal $y$ can be modeled as the superposition of the direct component and the IMS scattering component:

$$y = \left[ \sum_{l=1}^{L} \sqrt{P_R^l} (\alpha_l e^{j\omega_l}) e^{-jk(a_l + b_l)} + \sqrt{P_{T-R}} e^{-jkd_{T-R}} \right] x + noise \tag{S26}$$

where $d_{T-R}$ is the distance between Tx and Rx, and $P_{T-R} = \frac{\lambda^2}{(4\pi d_{T-R})^2}$ is the direct power received by Rx.

Addressing actual non-line-of-sight scenarios, assume there are $S$ equivalent scattering obstacles (blocks) between Tx and IMS, and assume the $s$-th block has a generalized Radar Cross Section (RCS) of $\sigma_{RCS}^s$. By separating the attenuation of the Tx-IMS and IMS-Rx links, the received power and discrete baseband signal evolve into:

$$P_R^{s,l} = \frac{G_e^2 \lambda^4 \sigma_{RCS}^s}{(4\pi)^5 a_s^2 b_{s,l}^2 c_l^2} \tag{S27}$$

Note that a generalized RCS does not imply that multiple block interactive paths will be treated as an interactive entity with an area; its power is still calculated according to the free-space transmission function between multiple points. Generalizing this model to the case of $S$ blocks and $L$ IMS elements, by redefining the channel model through separating the attenuation of the Tx-IMS and IMS-Rx links, the received discrete-time baseband signal can be expressed as:

$$y=\left[\sum_{l=1}^{L}\sqrt{Q_l^2}\,(\alpha_l e^{j\omega_l})e^{-jkc_l}\times\left(\sum_{s=1}^{S}\sqrt{Q_{s,l}^1}\,e^{-jk(a_s+b_{s,l})}\right)\right]x \tag{S28}$$

where the loss of the IMS-Rx link is $Q_l^2=\frac{G_e\lambda^2}{(4\pi c_l)^2}$, and the path loss of the Tx-blocks-IMS link is $Q_{s,l}^1=\frac{G_e\lambda^2\sigma_{RCS}^s}{(4\pi)^3 a_s^2 b_{s,l}^2}$, meaning $P_R^{s,l}=Q_{s,l}^1 Q_l^2$. By separating the link loss, the above equation can be rewritten in vector product form:

$$y=\boldsymbol{g}^{\boldsymbol{T}}\boldsymbol{\omega}\boldsymbol{h}x \tag{S29}$$

where $\boldsymbol{g}=\left[\sqrt{Q_1^2}e^{-jkc_1},\cdots\cdots,\sqrt{Q_L^2}e^{-jkc_L}\right]^T$ is the channel coefficient vector for the IMS-Rx link, $\boldsymbol{\omega}=diag([\alpha_1 e^{j\omega_1},\cdots\cdots,\alpha_L e^{j\omega_L}])$ is the IMS element response matrix, and $\boldsymbol{h}^{\boldsymbol{T}}=[\sum_{s=1}^{S}\sqrt{Q_{s,1}^1}\,e^{-jk(a_s+b_{s,1})},\cdots\cdots,\sum_{s=1}^{S}\sqrt{Q_{s,L}^1}\,e^{-jk(a_s+b_{s,L})}]$ is the channel coefficient vector for the Tx-blocks-IMS link. Notably, when the IMS is far from Tx and Rx, $b_{s,l}$ and $c_l$ can mathematically be assumed to be independent of the IMS element $U_l$, an analysis method similar to beam steering; in this case, the channel model is only related to the overall phase response of the IMS. Furthermore, considering the Tx-Rx link and noise, it can be extended to:

$$y=(\boldsymbol{g}^{\boldsymbol{T}}\boldsymbol{\omega}\boldsymbol{h}+\boldsymbol{h}_{\boldsymbol{LOS}}+\boldsymbol{h}_{\boldsymbol{blocks}})x+noise \tag{S30}$$

where $h_{LOS}$ represents the direct LOS link between Tx-Rx, and $h_{blocks}$ represents the non-line-of-sight link between Tx-Rx not regulated by IMS. For the LOS-dominant link, $h_{LOS}=\sqrt{P_{T-R}}e^{-jkd_{T-R}}$. For the NLOS link unadjusted by IMS, $h_{blocks}=\sum_{s=1}^{S}\sqrt{Q_s^1}\,e^{-jk(a_s)}\cdot\sqrt{Q_s^2}e^{-jk(d_s)}$, where $d_s$ represents the distance between Rx and the $s$-th block. This is equivalent to replacing the Tx-blocks-IMS path symmetrically with Tx-blocks-Rx, where $Q_s^1$ and $Q_s^2$ represent the losses from Tx and Rx to the blocks respectively.

Furthermore, this model can be generalized to complex indoor physical environments containing $T$ transmit antennas and $R$ receive antennas. At this time, the full-space deterministic MIMO channel matrix is modeled as follows:

$$\begin{bmatrix}y_1\\y_2\\\vdots\\y_R\end{bmatrix}=\left(\boldsymbol{G}_{\boldsymbol{1}}^{\boldsymbol{T}}\boldsymbol{\Omega}\boldsymbol{G}_{\boldsymbol{2}}+\boldsymbol{H}_{\boldsymbol{LOS}}+\boldsymbol{H}_{\boldsymbol{blocks}}\right)\begin{bmatrix}x_1\\x_2\\\vdots\\x_T\end{bmatrix}+\boldsymbol{noise} \tag{S31}$$

where $\boldsymbol{G}_{\boldsymbol{1}}$ and $\boldsymbol{G}_{\boldsymbol{2}}$ represent the losses from Tx and Rx after passing through the blocks to the IMS, respectively. $\boldsymbol{\Omega}$ is the coding pattern matrix, $\boldsymbol{H_{LOS}}$ is the $R \times T$ direct path matrix, $\boldsymbol{H_{blocks}}$ is the $R \times T$ blocked path channel matrix, and $\boldsymbol{noise}$ is the $R \times 1$ noise vector, all of which are generalized extensions of the original channel model.

In this full-space deterministic MIMO channel matrix, each component achieves a closed loop with the preceding geometric tracking and physical evaluation modules: the amplitude attenuation terms of the channel coefficient vectors (such as $\boldsymbol{g}, \boldsymbol{h}$, and various path losses $\boldsymbol{Q}$) are directly calculated by the electromagnetic physics proxy model in the second section; whereas the phase delay terms strictly depend on the precise physical distances output by the ray tracing engine in the first section.

Because this paper focuses on the regulation of a single transmitter beam by IMS at the simulation and actual measurement levels, only an IMS-assisted channel model for a single transmitter is used, but the aforementioned channel model offers an excellent reference for IMS-assisted channel modeling. In a deterministic indoor physical environment, $\boldsymbol{H_{LOS}}$ and $\boldsymbol{H_{blocks}}$ are approximately considered static, leaving only $\boldsymbol{\Omega}$ as a dynamic programmable variable. This modeling approach, which combines the determinism of ray tracing with the efficient expression of the channel matrix, not only outputs parameter-interpretable CSI in real-time but also provides complete physical layer data support for developing end-to-end communication simulations (such as OFDM systems) and accurately calculating channel capacity under dynamic control based on this twin environment in future work.

## Supplementary Note 5. Detailed Explanation of Online Rapid Optimization Algorithm for IMS Coding

Following the internal world model of channel prediction, the primary challenge transitions to how metaEI-WM can rapidly calculate the optimal RIS phase coding required to compensate for blind-zone fading. To address this, metaEI-WM employs a two-stage "prediction-inversion" rapid optimization algorithm. This algorithm seamlessly integrates physics-driven electromagnetic back-projection with modified computer-generated holography (CGH) techniques, enabling the highly efficient generation of the coding matrix and establishing an offline-optimized, perturbation-resistant closed-loop mechanism. The pseudocode of the algorithm is shown in

Algorithm 1.

---

Algorithm 1: Two-Stage Online Rapid Optimization for IMS Coding

---

Input: Geometry- and material-aware scene digital twins, transmitter/receiver coordinates, initial phase continuous matrix $\phi_0$, ideal amplitude $A_{ideal}$, max GS iterations $K_{max}$, RSSI threshold $\gamma_{th}$.

Output: Optimal 1-bit IMS coding matrix $C^*$.

// Stage 1: "Prediction - Inversion" Two-stage Rapid Optimization Algorithm

1: Extract $L$ dominant rays via ray tracing and spatial filtering in the geometry- and material-aware digital twin.

2: Initialize iteration index $k \leftarrow 0$.

3: While $k < K_{max}$ do

a) **Forward Propagation:** Compute the reconstructed field $E^{out(k)}(\boldsymbol{r})$ at the target using $\phi_k$ and dominant rays.

b) **Target Field Injection:** Enforce ideal point spread function amplitude:

$$\tilde{E}^{out(k)}(\boldsymbol{r}) \leftarrow |A_{ideal}| \cdot \exp\left(j \cdot \arg\left(E^{out(k)}(\boldsymbol{r})\right)\right)$$

c) **Backward Projection:** Back-propagate $\tilde{E}^{out(k)}(\boldsymbol{r})$ to IMS aperture using conjugate Green's function to obtain the optimal aperture radiation pattern $\tilde{E}^{surf(k)}(\boldsymbol{r}_{\boldsymbol{m},\boldsymbol{n}})$.

d) **Phase Decoupling:** Extract the reflection phase $\varphi_{m,n}^{ref} = \arg\left(\tilde{E}^{surf(k)}(\boldsymbol{r}_{\boldsymbol{m},\boldsymbol{n}})\right)$ and subtract the incident wave phase $\varphi_{m,n}^{in}$ obtained by ray tracing to get the phase shift $\Delta\varphi_{m,n} = \varphi_{m,n}^{ref} - \varphi_{m,n}^{in}$.

e) **1-bit Quantization:** Update discrete coding matrix $C^{k+1}$ : $C_{m,n}^{(k+1)} \leftarrow 1\ if\ \varphi_{m,n} \in \left[-\frac{\pi}{2}, \frac{\pi}{2}\right), else\ 0$..

f) $k \leftarrow k + 1$.

4: End while

5: Set initial physical matrix $C_{init} \leftarrow C^{K_{max}}$.

// Stage 2: Offline optimization & Spatial-Temporal Memorization.

6: Deploy $C_{init}$ to physical IMS and measure real environmental RSSI via mobile robot.

7: If $RSSI < \gamma_{th}$ then

a) Trigger local greedy search ($flip < 5\%$ critical atoms in $C_{init}$.

b) Evaluate real RSSI increments iteratively until convergence to peak RSSI.

c) Update optimal matrix $C^* \leftarrow Fine - turned\ matrix$.

8: Else:

a) $C^* \leftarrow C_{init}$.

9: End if.

10: Package $C^*$ with current 3D coordinates and semantic context.

11: Store into metaEI-WM's Spatial-Temporal Knowledge Graph for $O(1)$ future retrieval.

12: return $C^*$.

---

During the prediction stage, metaEI-WM utilizes the ray-tracing engine to initially launch probing ray beams into the geometric space encompassing the target blind zone. Subsequently, the system employs a spatial filtering mechanism to discard weak rays that have suffered significant power attenuation due to severe multiple absorptions or scatterings, retaining only the "effective dominant rays" that constitute the predominant energy contribution.

Subsequently, based on the time-reversal symmetry of electromagnetic fields, the algorithm initiates a back-projection procedure: it treats the target blind zone equivalently as a virtual "radiation source," and back-propagates the maximized signal field expected to be constructed in the blind zone along the extracted effective dominant ray paths back to the physical aperture plane of the IMS (as shown in Extended Data Fig. 4 (I)). This yields an optimal outgoing radiation pattern at the IMS aperture that can well cover the signal blind zone. Then, the Gerchberg-Saxton (GS) algorithm is employed to convert this optimal radiation pattern into the required reflection phase distribution on the IMS aperture plane, as shown in Extended Data Fig. 4 (II). This phase distribution is subsequently subtracted from the incident wave phase distribution obtained by the ray-tracing engine (Extended Data Fig. 4 (III)) to extract the specific phase shift required by the metasurface. This physical-layer operation effectively strips away the impact of complex channel fading, serving as a continuous theoretical baseline for subsequent 1-bit discretization.

Assuming the metasurface is located in the $z = 0$ plane, it is divided into $M \times N$

sub-wavelength atoms with physical dimensions $dx \times dy$, and the coordinates of the $(m,n)$-th atom are $r_{m,n}$. Its reflection complex coefficient is constrained by the 1-bit hardware state $C_{m,n} \in \{0,1\}$, i.e., $R_{m,n} = \exp(j\pi C_{m,n})$. Under the assumption simplifying the mutual coupling effect between adjacent elements, the forward-propagating electric field $E^{out}(\boldsymbol{r})$ reaching the target point $\boldsymbol{r}$ in the blind zone satisfies the discrete radiation summation model:

$$E^{out}(\boldsymbol{r}) = \sum_{l=1}^{L}\sum_{m=1}^{M}\sum_{n=1}^{N}\frac{\exp(-jk_0\|\boldsymbol{r}-\boldsymbol{r}_{m,n}\|)}{\|\boldsymbol{r}-\boldsymbol{r}_{m,n}\|}\cdot E^{in}(\boldsymbol{r}_{m,n})\cdot R_{m.n} \tag{S32}$$

Note that $|E^{out}(\boldsymbol{r})|$ here is not simply a "point-to-point" cumulative electric field, but the result of the combined action of multiple paths in the environment caused by ray tracing. $L$ represents the total number of "effective dominant rays", and $E^{in}(\boldsymbol{r}_{m,n})$ is also obtained by the ray-tracing engine by tracing the phase from Tx to each element on the IMS aperture plane, as shown in Extended Data Fig. 4 (III). To find the optimal $C_{m.n}$ combination, the system applies the modified Gerchberg-Saxton algorithm [9], mapping the continuous ideal phase to a 1-bit hardware-executable discrete matrix, ultimately generating a set of initial coding matrices suitable for the current physical environment. The specific flow of the algorithm is as follows:

(1) Forward field synthesis: In the $k$-th iteration, given the current continuous phase matrix $\phi^k$, calculate the holographically reconstructed complex electric field $E^{out(k)}(\boldsymbol{r})$ reaching the target plane.

(2) Spatial filtering and target amplitude injection: Extract the phase spectrum $\arg(E^{out(k)})$ of this reconstructed field, and forcefully replace its amplitude with the ideal point spread function $A_{ideal}$ (i.e., the impulse signal maximizing point focusing), thereby constructing the modified ideal target field $\widetilde{E}^{out(k)}$:

$$\widetilde{E}^{out(k)}(\boldsymbol{r}) = |A_{ideal}|\cdot\exp\left(j\cdot\arg\left(E^{out(k)}(\boldsymbol{r})\right)\right) \tag{S33}$$

(3) Back-projection: Treating the modified ideal field as a hologram, use the conjugate Green's function (i.e., reversing the sign of the propagation phase to $\exp(+jk_0 d)$ to back-propagate to the physical metasurface aperture plane, obtaining the desired modified aperture field distribution $\widetilde{E}^{surf(k)}(\boldsymbol{r}_{m,n})$.

(4) Phase Decoupling and 1-bit Nonlinear Projection: Extract the continuous reflection phase from

the back-projected aperture field, denoted as $\varphi_{m,n}^{ref} = \arg\left(\widetilde{E}^{surf(k)}(\boldsymbol{r}_{m,n})\right)$. To obtain the actual phase shift required by the metasurface, this reflection phase must be subtracted from the continuous incident wave phase $\varphi_{m,n}^{in}$ provided by the ray-tracing engine. Let the required continuous phase shift be $\Delta\varphi_{m,n} = \varphi_{m,n}^{ref} - \varphi_{m,n}^{in}$. Then, forcefully map this continuous phase difference to the discrete hardware physical constraint space to update the coding matrix:

$$C_{m,n}^{(k+1)} = \begin{cases} 1, \ \varphi_{m,n} \in [-\frac{\pi}{2}, \frac{\pi}{2}) \\ 0, \ \varphi_{m,n} \in [-\pi, -\frac{\pi}{2}) \cup [\frac{\pi}{2}, \pi) \end{cases} \tag{S34}$$

(5) Error evaluation and termination criterion: Benefiting from the energy-conserving convergence characteristics, this alternating projection process typically enters a convergence plateau within 15-20 iterations, finally outputting the initial optimal 1-bit coding matrix suitable for the current physical environment. Extended Data Fig. 4 (IV) and (V) respectively show the optimized coding matrix and the corresponding far-field outgoing radiation pattern generated in the example.

Leveraging the precise spatial reconstruction afforded by the high-fidelity digital twin model and the efficient optimization capabilities of the modified GS algorithm, the initial coding matrix provides substantial coverage gain in the vast majority of scenarios. Nevertheless, under extreme conditions, significant unmodeled dynamic obstacles (e.g., sudden object movements) or accumulated computational errors within the model may cause the measured RSSI to fall below the requisite communication quality threshold. Consequently, relying exclusively on simulation-generated coding may result in performance degradation within the real physical environment, thereby necessitating the activation of a subsequent offline fine-tuning closed-loop mechanism.

Under the offline fine-tuning mechanism, triggered either by an explicit user request or when the system is in an idle state without ongoing tasks, the intelligent mobile vehicle autonomously navigates to each target grid requiring signal enhancement to conduct on-site signal acquisition; synchronously, the metasurface iteratively updates its corresponding coding matrix. This process utilizes the phase matrix output by the modified GS algorithm as a high-quality initial value and employs a local greedy search strategy with exceptionally low computational complexity.

Specifically, the algorithm randomly flips the binary coding states (0/1) of a marginal fraction (typically less than 10%) of key meta-atoms on the metasurface, and evaluates the performance using the real-time empirical signal quality collected by the vehicle as feedback. If the signal quality improves, the state change is retained; otherwise, the coding matrix reverts to its previous iteration. Because this exploration is confined to the neighborhood of a near-global optimum, the closed-loop process typically requires only a minimal number of sampling steps to maximize the received power of the physical channel, thereby effectively compensating for the theoretical modeling discrepancies inherent in the geometry- and material-aware digital twin.

Upon completion of the closed-loop offline fine-tuning, metaEI-WM not only restores network coverage within the blind zone but also structurally integrates the converged optimal coding matrix with its corresponding 3D physical coordinates and surrounding architectural semantic context, subsequently archiving this data within the agent's spatiotemporal knowledge graph. Throughout extended operational cycles, when a communication node re-enters the blind zone or the environment exhibits a similar semantic topological distribution, the system can directly retrieve and deploy the historical optimal matrix configuration from the knowledge graph. This "compute-once, benefit-permanently" memory mechanism effectively reduces the average time complexity for coding generation to the $O(1)$ level, demonstrating exceptional scenario adaptability and minimal computational overhead.

## Supplementary Note 6. The System Implementation of metaEI-WM for Spatial Perception and Non-Contact Physiological Monitoring

This supplementary document aims to expound upon the underlying hardware-software co-design mechanisms, mathematical derivations, and algorithmic paradigms employed by metaEI-WM during the validation of its dynamic signal enhancement and spatial sensing capabilities for mobile users. The methodologies encompassed herein include high-precision spatial coordinate reconstruction, multi-level human skeletal topology extraction, the generalized Gerchberg-Saxton (GS) near-field focusing algorithm, and a Variational Mode Decomposition (VMD)-based extraction mechanism for weak respiratory signals. Collectively, these components comprehensively substantiate the exceptional performance and scientific significance of metaEI-WM in bridging digital semantic comprehension with physical-layer beam manipulation (see Fig. S4).

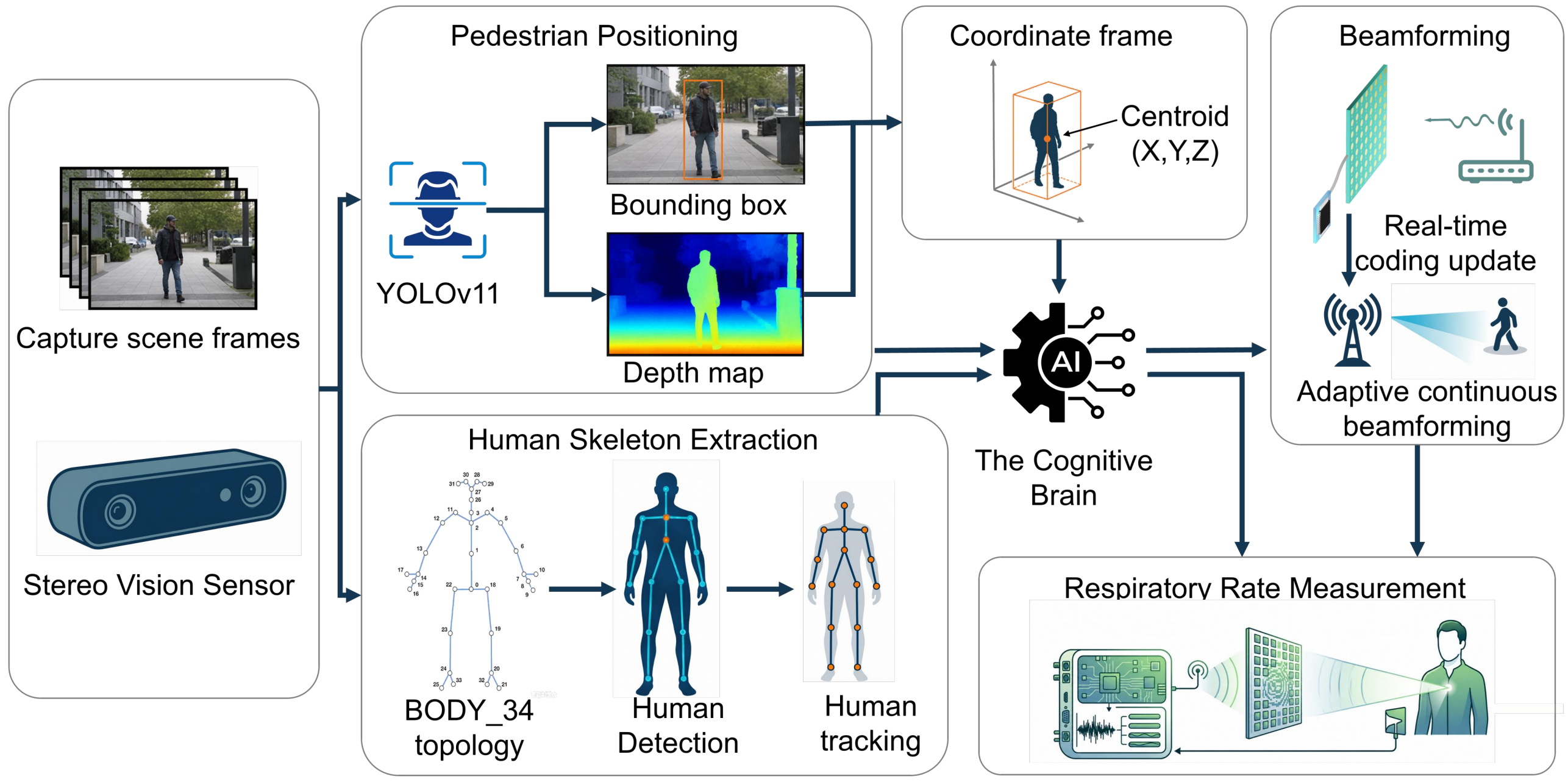

**Fig. S4. The system implementation of metaEI-WM for spatial perception and non-contact physiological monitoring .** The system integrates a ZED 2i depth camera to capture scene frames. In the pedestrian positioning module, the lightweight YOLO11 framework and a 32-bit depth map are combined to extract the absolute 3D spatial coordinates (X, Y, Z) of the pedestrian's centroid. For refined non-contact health monitoring tasks, the system utilizes the BODY_34 skeletal topology for 2D/3D human detection and 3D human tracking. The "cognitive brain" utilizes these coordinates to guide real-time updates to the phase properties of metasurface atoms. This achieves adaptive and continuous beam deflection for robust communication links , and facilitates respiratory rate measurement through a weak respiratory signal extraction mechanism based on the Variational Mode Decomposition (VMD) algorithm.

### 1. Channel Link Enhancement under Dynamic Target Tracking

In highly dynamic communication environments, user mobility is prone to inducing severe multipath effects and channel fading, leading to a significant degradation in communication link quality, which in turn impairs data throughput and connection stability. To fundamentally resolve this physical-layer challenge, the metaEI-WM employs an adaptive continuous tracking protocol based on visual-inertial multi-sensor fusion. This protocol provides the "cognitive brain" with the absolute 3D spatial coordinates of the target with extremely low latency, thereby guiding real-time updates to the phase properties of the metasurface atoms to achieve adaptive and continuous beam deflection[10].

To achieve low-latency, high-precision extraction of pedestrian spatial coordinates in complex physical environments, the metaEI-WM integrates an in-situ Stereolabs ZED 2i depth camera. In the spatial localization mode, the metaEI-WM introduces the latest generation YOLO11 object detection framework[11] to execute low-latency, 2D image-level pedestrian recognition and bounding box extraction. In complex indoor scenarios, targets may be partially occluded by furniture or located in extreme backlighting or low-light conditions. To address these challenges, YOLO11 incorporates multiple deep innovations in its network architecture, significantly enhancing the robustness of its feature extraction. It substantially reduces its total parameter count by 22% while maintaining or even exceeding an extremely high mean Average Precision (mAP). This lightweight network characteristic directly translates into exceptionally high inference frame rates, fully satisfying the stringent requirements of dynamic electromagnetic beam modulation for millisecond-level system response latency.

The engineering implementation of joint 3D real-time coordinate extraction by YOLO11 and ZED 2i relies on the highly optimized TensorRT framework within the ZED SDK. Its workflow is as follows: first, the ZED 2i's high-resolution binocular RGB sensors capture synchronized video frames. Subsequently, the exported YOLO11 ONNX model is natively loaded into the CUSTOM_ YOLOLIKE_BOX_OBJECTS mode of the ZED SDK API for low-level hardware-accelerated inference. Concurrently with YOLO11 outputting 2D bounding boxes, the neural depth engine inside the ZED SDK synchronously generates a high-precision 32-bit depth map containing scene depth variations. This depth map precisely records the absolute metric distance from behind the left lens of the camera to the physical objects within the scene. Based on this, the system computes a 2D mask to determine the pixel set belonging to the pedestrian target and maps it to the corresponding location on the depth map. This allows the precise projection of the 2D detection box into a 3D bounding box, thereby extracting the absolute 3D spatial coordinates (X, Y, Z) of the pedestrian's centroid. Combined with the high-frequency IMU data and stereo visual odometry of the ZED 2i, the system's Visual-Inertial Simultaneous Localization and Mapping (VSLAM) algorithm ensures that the coordinate data remains absolutely accurate within the global coordinate system, even in the presence of slight camera displacements.

After obtaining the user's real-time 3D spatial coordinates, the core task of the metaEI-WM is to rapidly derive the optimal metasurface phase coding matrix via its internal electromagnetic dynamics prediction and decision execution module. To completely eliminate the substantial system latency caused by the online real-time calculation of complex Green's functions or ray tracing and to ensure real-time responsiveness, the metaEI-WM adopts a strategy of "offline pre-computation and online dictionary retrieval". Specifically, the system partitions the target observation area into a finite number of discrete spatial nodes. During system idle periods, a closed-loop calibration algorithm pre-calculates the optimal phase coding configuration for each spatial grid node by combining the internal physical world model with a ray-tracing engine. This pre-calibrated mechanism establishes a vast and precise "coordinate-coding mapping dictionary," enabling the system, while in an active running state, to instantaneously output the optimal matrix simply by executing a sub-microsecond table lookup operation.

To quantitatively evaluate and verify the physical effectiveness of this dynamic signal enhancement strategy, the system rigorously models and measures the empirical channel capacity of mobile users along specific trajectories mathematically using the Shannon-Hartley theorem. For a human user moving along a predetermined trajectory, the channel capacity when located at the $n$-th spatial node is defined as:

$$C_{persion} = Blog_2\left(1 + \frac{S(n)}{N(n)}\right) \tag{S35}$$

where B represents the radio frequency spectrum bandwidth occupied by the USRP transceiver system, $S(n)$ denotes the effective signal power captured at the $n$-th node, and $N(n)$ represents the background thermal noise present at the $n$-th node. Combined with the measurement results in **Fig. 4d**, it is evident that the system's average channel capacity achieves a substantial and consistent improvement.

**2. Human Skeletal Topology Extraction and Near-Field Beam Focusing**

While the aforementioned 2D bounding box extraction and 3D centroid localization mechanisms are sufficient to meet the enhancement requirements of macro communication channels, a single centroid coordinate proves too coarse when facing more refined non-contact health monitoring tasks, such as respiration detection. At this juncture, the cognitive brain of the metaEI-WM triggers a smooth algorithmic switching mechanism, transitioning seamlessly from the YOLO11 engine used for rigid bounding box extraction to the ZED 2i's native advanced human skeleton detection engine. This skeleton-tracking mechanism is driven by a deep neural network specifically optimized for human kinematics, capable of outputting extremely detailed biomechanical data. To ensure the accuracy of human joint coordinates and handle occlusion issues, the advanced skeleton tracking process is rigorously divided into three logical levels:

- **2D/3D Human Detection:** The neural network first performs deep feature extraction on high-resolution RGB images to infer the 2D planar pixel coordinates of key human joints. Subsequently, the system injects these 2D coordinate arrays into the depth computation and position tracking modules, utilizing the high-fidelity depth map to directly resolve the absolute coordinates of each skeletal joint in 3D space.
- **3D Human Tracking:** When processing video stream data, the system utilizes a multi-object tracking algorithm to assign consistent identity labels to the same detected human body across multiple consecutive frames. Simultaneously, by calculating the 3D displacement differences of each joint between adjacent frames, the system can derive the velocity vectors of individual limb components, providing a dynamic basis for subsequent action intention recognition.
- **3D Body Fitting:** This is the critical step to ensure that the skeletal structure conforms to biological logic. Due to the restricted field of view of the depth camera, some limbs will inevitably be occluded by the torso or other objects. The system utilizes prior knowledge of human kinematics to accurately and completely reconstruct the full-body posture, even with partial occlusion, by calculating the local rotation tensors between adjacent bones and performing inverse kinematics calculations in 3D space.

In the specific configuration of the metaEI-WM, the cognitive brain adopts the highly refined BODY_34 skeletal topology format, which can extract 34 standard anatomical keypoints distributed across the human body in real-time. These 34 nodes not only cover the major joints of the limbs (such as shoulders, elbows, wrists, hips, knees, and ankles) but, more importantly, deeply characterize the directional distribution of the spinal central nervous system and facial orientation features.

In subsequent vital sign sensing tasks, such as respiration monitoring, micrometer-level chest undulation is the sole data source. Therefore, precisely locking onto the absolute spatial position of the chest cavity is the paramount priority of the entire perception link. In the BODY_34 index library, the CHEST_SPINE (keypoint index 2) serves as the system's most crucial physiological targeting anchor. This coordinate is input into the electromagnetic beam focusing module as the absolute target center for energy concentration, ensuring that the backscattered signals contain the maximum chest volume variation features.

To extract faint respiratory frequencies from mobile users in a non-contact manner, the metaEI-WM must overcome the issue of extreme signal attenuation. Human respiratory activity physically manifests as periodic micro-displacements of the chest wall moving back and forth by a few millimeters to a centimeter at a specific low frequency (typically 0.2 Hz to 0.4 Hz). This displacement imposes an extremely weak phase modulation on incident electromagnetic waves. If an omnidirectional antenna is used to blindly receive environmental reflections, this weak physiological phase signature will be completely drowned in clutter. Therefore, the system implements extreme near-field beam focusing through a high degree of hardware-software collaboration.

In the experiments conducted in the Warehouse space mentioned in the main text, the system abandons traditional integrated radar architectures in favor of a more flexible software-defined radio platform. The core control module is the USRP B210, which allows the system to define the timing, frequency bands, and waveforms of transmission and reception with programming-level precision. The transmitting end employs an omnidirectional antenna to radiate radio frequency energy evenly into space and illuminate the metasurface panel. The receiving end is equipped with a Vivaldi patch antenna. Thanks to its excellent ultra-wideband characteristics, high directivity, and smooth phase response, the Vivaldi antenna serves as the ideal sensor to capture backscattered microwave echoes modulated by the chest cavity. The metasurface's operating frequency is set to 3.5 GHz; its wavelength can easily penetrate common winter coats or shirts, forming effective reflections and scattering directly on the surface of the human chest wall, guaranteeing the physical integrity of the backscattered signal. When the user is within the unoccluded near-field region of the metasurface, the wavefront exhibits distinct spherical wave characteristics, rendering traditional phased-array far-field beamforming formulas based on plane wave approximation completely invalid here. To form a sharp electromagnetic energy focal point at any specific location in 3D space (i.e., the CHEST_SPINE 3D coordinate extracted by BODY_34), the metaEI-WM utilizes the near-field Gerchberg-Saxton algorithm. By defining an ideal intensity distribution at the target user's chest location, the algorithm inversely derives and calculates the optimal phase coding matrix required on the holographic metasurface transmitting face. Through iterative GS calculations, the metasurface dynamically generates a holographic phase map within hundreds of microseconds, converging radio frequency energy onto the user's undulating chest and returning a physiological reflected wave with an extremely high signal-to-noise ratio to the receiving end.

## 3. Non-Contact Weak Respiration Signal Extraction Mechanism Driven by Variational Mode Decomposition

Despite physical-layer beam focusing via the GS algorithm, the baseband signal captured by the Vivaldi receiving antenna remains an extremely complex mixed time-domain waveform. The human body is a dynamic and complex scatterer; the echo not only contains the slow periodic displacement of the chest caused by breathing (0.1~0.5 Hz frequency band), but is also mixed with much weaker high-frequency tremors caused by heartbeats, alongside inevitable spatial white noise and low-frequency baseline drift induced by slight body swaying. To accurately resolve a pure respiration feature curve from this chaotic mixed signal, the metaEI-WM deploys a rigorous advanced signal processing pipeline. The system first applies routine anti-aliasing preprocessing and frequency band isolation to the received signal to effectively filter out ultra-low-frequency noise below the breathing frequency, as well as high-frequency thermal noise, mains interference, and electromagnetic noise from other equipment in the environment. Subsequently, the system adopts the more advanced and highly robust VMD algorithm for respiration detection[12].

VMD is a fully non-recursive signal decomposition architecture capable of concurrently extracting multiple modes of a signal within a unified framework. The core concept of VMD is to decompose the input signal $f(t)$ into $K$ discrete Intrinsic Mode Functions (IMFs), denoted as $u_k(t)$, while requiring that each mode oscillates closely around its corresponding central frequency $\omega_k$.

The VMD algorithm reformulates the signal decomposition problem into a strictly constrained variational optimization problem, whose mathematical objective is: to find these $K$ modes and their central frequencies such that the sum of the estimated bandwidths of each mode is minimized, and the sum of these $K$ modes must perfectly reconstruct the original signal. The algorithm performs narrow-band prior calculations in the complex frequency domain, and the physical steps to compute the bandwidth include:

(1) For each mode $u_k(t)$, its analytical signal is constructed via the Hilbert transform, thereby obtaining a single-sided frequency spectrum.

(2) By multiplying the analytical signal by an exponential term $e^{-j\omega_k t}$, the frequency spectrum of the mode is shifted to the baseband, aligning it with its central frequency position.

(3) The demodulated signal is differentiated, and the squared $L_2$-norm of this gradient is calculated to rigorously estimate the bandwidth.

The rigorously constructed variational constraint formulation is as follows:

$$\min_{\{u_k\},\{\omega_k\}} \{ \sum_{k=1}^{K} \left\| \partial_t \left[ \left( \delta(t) + \frac{j}{\pi t} \right) * u_k(t) \right] e^{-j\omega_k t} \right\|_2^2 \} \tag{S36}$$

Subject to the constraint: $\sum_{k=1}^{K} u_k(t) = f(t)$. To solve the highly challenging constrained optimization problem, the system introduces a quadratic penalty factor and Lagrangian multipliers. These are used to accelerate convergence, improve reconstruction accuracy in high-variance noise environments, and strictly enforce data fidelity constraints, thereby transforming it into an unconstrained optimization problem. Finally, iterative optimization is performed using the Alternating Direction Method of Multipliers (ADMM) until the central frequencies and modes are completely separated and converged.

Through the highly efficient decomposition by VMD, the original mixed signal is perfectly dissected into multiple IMF components with mutually non-interfering frequency bands. Given that the respiratory frequency of a normal adult in a resting or mildly active state typically ranges from 12 to 20 breaths per minute, the corresponding central frequency $\omega_k$ falls within the 0.2 Hz to 0.33 Hz range. The system directly extracts the single IMF whose central frequency lies within this physiological interval from the decomposed modes, utilizing it as an exceptionally pure, non-contact lung respiration feature curve.

However, merely obtaining the time-domain waveform does not equate to acquiring specific respiration rate values. Because human breathing is not a perfectly absolute periodic sine wave, occasional respiratory pauses or deep breaths can cause local waveform distortions, making traditional simple zero-crossing detection highly susceptible to severe misjudgments. Therefore, the final link in the algorithm is the deployment of an adaptive dynamic peak-finding mechanism that combines peak and trough detection. This mechanism is executed within a sliding time-domain window:

(1) **Signal Normalization and Extrema Calibration:** It first identifies all local maxima (peaks) and minima within the window.

(2) **Dynamic Amplitude Threshold Cleaning:** A statistical amplitude threshold is set (for example: rejecting minor spikes with an amplitude lower than 0.2 times the average amplitude of the top 75% peaks within the window) to forcibly eliminate pseudo-peaks generated by high-frequency glitches or micro-tremors.

(3) **Temporal Physiological Gating Constraint:** Based on human physiological limits, there is an absolute minimum time interval between two consecutive inhalations. The algorithm forcibly clears any extrema point that is less than 0.5 seconds away from the previous valid peak, physically precluding over-counting.

By calculating the time difference between two adjacent valid peaks, the system can resolve a continuous, high-resolution dynamic respiration rate curve in real-time. As demonstrated by the comparative experimental results in the main text, when this non-contact extracted curve is aligned and evaluated against the data output by a medical-grade contact reference device (ADS1292R heartbeat detection module) worn by the user, the two exhibit an extremely high degree of consistency, proving the engineering reliability of this signal decomposition mechanism centered on VMD.